\documentclass[12pt]{amsart}          %
\usepackage{amsfonts} 
\usepackage{epsfig}
\usepackage{hhline,eufrak,euscript,delarray}

\newtheorem{theorem}{Theorem}[section]
\newtheorem{lemma}[theorem]{Lemma}
\newtheorem{corollary}[theorem]{Corollary}
\newtheorem{rem}[theorem]{Remark}

\author{ Vlady RAVELOMANANA }
\email{vlad@lipn.univ-paris13.fr}
\address{Vlady RAVELOMANANA, LIPN -- UMR 7030, Institut Galil\'ee --Universit\'e de Paris-Nord,
99, Avenue J. B. Cl\'ement. F 93430 Villetaneuse, France.}
\title[]{The Average Size of Giant Components Between the Double-Jump}

\def \equaldef{\stackrel{\tiny{\mbox{def}}\normalsize}{=}}
\def\kp{{k_{\ast}}}
\def\xp{{x_{\ast}}}
\def\qE{{\mathbb{E}}}
\def\qP{{\mathbb{P}}}
\def\a{\rho}
\def\V{\vartheta}

\def \beq{\begin{equation}}
\def \eeq{\end{equation}}
\def \be{\begin{eqnarray*}}
\def \ee{\end{eqnarray*}}
\def \ben{\begin{eqnarray}}
\def \een{\end{eqnarray}}
\def \pl{\medskip}

\def \B{\Big}
\def \b{\big}

\def \l{\left}
\def \r{\right}
\def\coeff#1{\left[ #1 \right]}
\def\point{\noindent $\, \, \, {\mathbf{\bullet}}\, \, \,$}
\def\EIS#1{EIS~ {\bf #1}}

\def\auteur{}

\newcommand{\ENDPROOF}{\hfill \qed}

\def \ten{\rightarrow}
\def\Gt{$  \{ {\mathbb{G}}(n,t)\} _{0 \leq t\leq 1} \,$}
\def\Gp{$  \{ {\mathbb{G}}(n,p)\}$}

\def\VL{{V_{n}^{(\ell)}}}

\def\Err{\nu}

\def\dist{\stackrel{\tiny{\mathbf{d}}\normalsize}{\ten}}
\def\qE{{\mathbb{E}}}

\begin{document}
\label{firstpage}

\maketitle

\begin{abstract}
We study
 the sizes of  connected components according to their
 excesses 
during a random graph process built with $n$ vertices. 
The considered model is the \textit{continuous} one defined in \cite{Ja2000}. 
An ${\ell}$-component is
 a connected component with ${\ell}$ edges more than vertices. 
$\ell$ is also called the \textit{excess} of such component.
As our main result, we show that when $\ell$ and ${n \over \ell}$ are both large, the 
expected number of vertices that ever belong to an $\ell$-component
is about ${12}^{1/3} {\ell}^{1/3} n^{2/3}$. We also obtain limit 
theorems for the number of creations of $\ell$-components.
\end{abstract}


\section{ \bf Introduction}

Following Erd\H{o}s and  R\'enyi's  pioneering
 works around 1960 \cite{ER59, ER60}, random graphs have been the
subject of intense studies for four decades.
Topics on random graphs provide a large and particularly active body of
research. We refer to  the books of Bollob{\'a}s \cite{Bollobas},
of Kolchin \cite{Kolchin},  and of Janson, \L uczak and Ruci\'nski 
\cite{JLR2000} for excellent treatises related to these subjects.

We consider here labelled graphs on  vertex set $V=\{1,2, \ldots, n\}$ with 
undirected edges without self-loops or multiple edges.
The set of all such graphs is
denoted by ${\mathcal{G}}^n$ and, a random graph is
defined by  a pair $({\mathcal{G}}^n, P)$ where $P$ is
a probability distribution over ${\mathcal{G}}^n$.
Let us recall the three  popular processes of random graphs
in the literature.
The first one, $\{ {\mathbb{G}}(n,M)
 \}_{0 \leq M \leq {n \choose 2} }$, consists of all graphs with vertex
set $V=\{1,2, \ldots, n\}$ having $M$ edges, in
which one can randomly pick a  graph with the same probability.
Thus, with $N={n \choose 2}$, we have $0 \leq M \leq N$
and the random graph ${\mathbb{G}}(n,M)$ has ${N \choose M}$ elements with
each element occurring with probability ${N \choose M}^{-1}$.
Secondly, $\{ {\mathbb{G}}(n,p) \}_{0 \leq p \leq 1}$ ,
 consists of all graphs with
the same vertex set $V=\{1,2, \ldots, n\}$ in which each of
the $N$ edges is  drawn independently with probability $p$.
The third process, \Gt (cf. \cite{Ja93,Ja2000}), may be constructed by letting  each edge $e$,
chosen amongst the $N$ possible edges, appear at
random time $T_e$, where $T_e$  are independent
random variables uniformly distributed on $[0,1]$. 
The random graph ${\mathbb{G}(n,t)}$
is constructed with all edges $e$ such that $T_e \leq t$.
The main difference between $\{{\mathbb{G}}(n,M)\}_{0 \leq M \leq {n \choose 2}}$ and \Gt is that
in  the first one, edges are added at fixed (slotted)
times $1, 2, \ldots$, $N$ so at any time $T$ we obtain
a random graph with $n$ vertices and $T$ edges,
whereas in \Gt the edges are added at random times.
 At time $t=0$, we have a graph with $n$ vertices and $0$ edges,
and as the time advances all edges $e$ with r.v.~ $T_e$
such that  $T_e \leq t$ (where $t$ is the current time),
 are added to the graph until $t$ reaches $1$
in which case, one obtains the complete graph $K_n$.

Following our predecessors \cite{Ja93,Ja2000,JKLP93,PITTEL-WORMALD}, 
let us define the {\it excess} or the {\it complexity} of a
connected graph as the difference between its number of edges
and its number of vertices. Throughout this paper,
as the random graph process proceeds, we will often
{\it fix} and study an arbitrary chosen  connected component
built with  $k \leq n$ vertices
(where $n$ is the total number of vertices) in the graph.
For ${\ell} \geq -1$, a $(k,k+{\ell})$ connected graph
is one having $k$ vertices and $k+{\ell}$ edges, thus its excess
is ${\ell}$ and we simply called it an
{\it ${\ell}$-component}.
A random graph process begins with a set of $n$ isolated vertices.
Then, as evolution proceeds, edges are added at random
(drawn without replacement) and at first, all components created are
trees ($(-1)$-components), later  $0$-components (also
called {\it unicyclic} components) will
appear and eventually the first ${\ell}$-components are created, with ${\ell} > 0$.
Usually, ${\ell}$-components are called {\it complex} whenever ${\ell} > 0$. 

As more edges are added, a complex component gradually
swallows up some other ``simpler'' components, and it is 
worth-noting that with nonzero probability, at least two components 
can co-exist as the random graph evolves \cite{Ja93,JKLP93}. We denote by $\VL$  the number of
 vertices that at some stage of the random graph process
belong to an ${\ell}$-component.

In this paper, we consider
the \textit{continuous time} random graph process \Gt,
and  we will study the creation of $({\ell}+1)$-components (${\ell} \geq 0$).
We can observe that there are two manners to create a
new $({\ell}+1)$-component during the random graph process~: \\
\point either by adding an edge inside  an existing ${\ell}$-component,\\
\point or by joining with the last added edge
a $p$-component to an $({\ell}-p)$-component, with $p \geq 0$.\\
Following Janson's notations \cite{Ja2000}, the first transition will be denoted
$\ell \rightarrow \ell+1$ and the second one $(\ell-p) \oplus p \rightarrow \ell+1$.\\
\noindent We  study the random variable $X_n^{({\ell})}$,
defined as the number of creations of $({\ell}+1)$-components
 during the evolution of the random graph.
As in \cite{Ja93},  denote respectively, by $Y_n^{({\ell})}$ and $Z_n^{({\ell})}$
the number of $({\ell}+1)$-components created by the two ways described above.
More precisely, $Y_n^{({\ell})}$ equals the number of
edges  added inside an ${\ell}$-component
creating an $({\ell}+1)$-component, and $Z_n^{({\ell})}$ is the 
number of bridges added  between a
$p$-component and an $({\ell}-p)$-component, for all $0 \leq p \leq \ell$ during the evolution
of the graph. Thus, by construction $X_n^{({\ell})} = Y_n^{({\ell})} + Z_n^{({\ell})}$.

\subsection{ \bf Related works}

In a former paper, Janson \cite{Ja93} obtained limit theorems
for the number of {\it complex components}, i.e., components with more
than one cycle, created during the evolution of the graph. In particular, 
Janson   computed  the probability that the process never contains
more than one complex component is approximately 0.87 (as the number of
vertices $n$ tends to infinity). Thus, at least two complex
components can co-exist in the random graph and  there is 
not a zero-one law for this process.
With the notations of our paper, Janson obtained limit laws
for $X_n^{({\ell})}, Y_n^{({\ell})}$ and $Z_n^{({\ell})}$ for ${\ell}=1$ (see for instance
\cite{Ja93} for precise statements of his results).
Using enumerative and analytical methods, Janson, Knuth, Pittel
and \L uczak \cite{JKLP93} obtained also the exact value
$5 \pi/18 = 0.872\cdots$
for the limit described above.

In \cite{BCM90, BCM92}, Bender
{\it et al.} studied  several properties of labelled graphs.
They computed the asymptotic number of connected graphs with $k$ vertices and
$k+{\ell}(k)$ edges for every function ${\ell}(k)$ as $k \ten \infty$.
Define a {\it bridge} or a {\it cut edge} of a connected 
component as an edge whose deletion will deconnect
the graph. Working in the probability space of connected components,
Bender, Canfield and McKay also  obtained the asymptotic probability
for a random chosen edge to be a bridge. 
 See for instance \cite[Section 5]{BCM92}.
 
Speaking about the largest component in $\mathbb{G}(n,M)$, 
Erd\H{o}s and R\'enyi \cite{ER60} suggested that a ``\textit{double jump}'' occurs:
the largest component changes its size (with respect to the number of vertices $n$) twice -- from $O(\log{n})$
to $O_p(n^{2/3})$ -- 
and then from $O_p(n^{2/3})$ to $O(n)$. Note that we use here the notation
$X_n = O_p(a_n)$ (e.g. \cite[p. 10]{JLR2000}):  
For a r.v. $X_n$ and real positive numbers $a_n$,
we have $X_n = O_p(a_n)$ as $n \ten \infty$ if $\forall \delta>0$
there exist constants $c_\delta$ and $n_0$ s.t.~ 
 $\qP(\|X_n\| \leq c_\delta a_n) > 1 - \delta$, for $n\geq n_0$. In particular,
Erd\H{o}s and R\'enyi expected that whatever function $M \equiv M(n)$ we choose, the 
largest component of $\mathbb{G}(n,M)$ can only be either
$O(\log{n})$ or $O_p(n^{2/3})$ or $O(n)$. In the latter case
 and for the Bernoulli random graph $\mathbb{G}(n,p)$, for $p=c/n$ with $c>1$, 
Barraez, Boucheron and De la Vega \cite{LALO}
have studied precisely the size of the giant component.
We refer also to \cite{LDP} where, among other results, 
O'Connell has investigated the size
of the giant component by means of large deviation principles.
Therefore, under the Bernoulli model, 
it is known that for $p=c/n$ with $c>1$, the size of the largest 
connected component, denoted $V_n$ is asymptotically $a n$,
where $a>0$ satisfies $a=1-e^{-a c}$ and the
sequence $V_n/n$ converges in probability to $a$.
Bender, Canfield and McKay \cite{BCM92} have also determined the probability that
a random graph produced under the \Gp~ process is connected as well as 
the asymptotic distribution of the number of edges of such a graph (conditioned on connectedness).
Pittel and Wormald \cite{PITTEL-WORMALD} presented an alternative \textit{inside-out} 
approach based on the
enumeration of graphs of minimum degree $2$. In particular, they obtained 
the asymptotic number of connected graphs with $n$ vertices 
and $M$ edges \cite[Theorem 3]{PITTEL-WORMALD},
as well as the joint limiting distribution of the \textit{size} of 
the \textit{$2$-core} (number of vertices of degree
at least $2$) of the giant component, its
\textit{excess}
 (number of edges minus number of vertices) and the size of its \textit{tree mantle} 
 (number of vertices of the giant component not in its $2$-core). 
Their results hold for the two models
of random graphs $\mathbb{G}(n,p)$
 and  ${\mathbb{G}}(n,M)$ in the so-called \textit{supercritical case}, i.e.,
 when $n^{1/3} (np - 1) \ten \infty$ or $n^{1/3} (2M/n - 1) \ten \infty$.

\subsection{ \bf Our results} \label{OUR_RESULTS}
The kind of problems discussed here
are in essence combinatorial. And as already noticed by Janson in \cite{Ja2000},
combinatorics and probability theory are closely related in such a way
that the combination of both approaches can help to study
the extremal characteristics of  indecomposable structures
typified by  random graphs. 

In order to study the random variables $X_n^{({\ell})}, \, Y_n^{({\ell})}$ and 
$Z_n^{({\ell})}$, we will use the method of moments (e.g. \cite[page 144]{JLR2000}). We will
investigate the factorial moments $\qE(Y_n^{({\ell})})_m$ (resp. $\qE(Z_n^{({\ell})})_m$)
 starting with the simplest cases, viz. the expectations. We will rely 
 $Y_n^{({\ell})}$ and $Z_n^{({\ell})}$ by means of enumerative/analytic tools
such as those developed in \cite{FKP89,JKLP93} and in \cite{BCM90,BCM92}.
First, we observe that $(Y_n^{({\ell})})_m$ is the number of ordered $m$-tuples
of edges that are added to $\ell$-components (both ends of the edges are in the components)
during the evolution of the random graph process. Similarly,
$(Z_n^{({\ell})})_m$ is the number of ordered $m$-tuples of edges that are added between
 pairs of disjoint complex components to build an $(\ell+1)$-component. 
As we shall see $(Y_n^{({\ell})})_m$ can be deduced using asymptotic results namely from
  \cite{BCM90} and \cite{PITTEL-WORMALD}. Therefore,
 our first task is to quantify the number of manners to build an $(\ell+1)$-component
 arising from the second type of transition.

More precisely, 
for the \textit{Wright's range}, i.e. for connected components built
with $k$ vertices and $k+o(k^{1/3})$ edges (this is the same range as in \cite{Wr80}),
we will  use the analytical tools
associated to the generating functions of Cayley's rooted trees \cite{Cayley},
$T(z)$, which plays an important role in the enumerative point of view
of  the general theory of random graphs (cf. the ``giant paper'' \cite{JKLP93}).
Next, for excesses greater then $o(k^{1/3})$, 
 we will use the results of
Bender, Canfield and McKay in \cite{BCM90}.


\noindent
 $\mathbf{\bullet}$~ 
 As a first result, we obtain Theorem \ref{THM_WAYS} which is
closely related to the r.v. $Z_n^{(\ell)}$ defined above.~ 
 We prove that~ {\it almost all}~ $({\ell}+1)$-components
whose {\it last added edge} forms a bridge (or a cut edge)
between a $p$-component and an $({\ell}-p)$-component, for
$0\leq p \leq {\ell}$, are built by linking a unicyclic
component to an ${\ell}$-component. In fact, we have the following theorem:
\begin{theorem} \label{THM_WAYS}
Denote by $c(r,s)$ the number of connected graphs with $r$ vertices and
$s$ edges. 
As $k,{\ell} \ten \infty$ and ${\ell}  \ll k$ 
the number of ways, $c'(k,k+{\ell}+1)$, to build an
$({\ell}+1)$-component of order $k$ with a distinguished cut edge
between a $p$-component and an $({\ell}-p)$-component, $p\geq 0$,
 satisfies
\ben
c'(k,k+{\ell}+1)
& = & \frac{1}{2}  \sum_{p=0}^{\ell} \sum_{t=1}^{k-1}
{k \choose t} t(k-t) \, c(t,t+p) \, \,  c(k-t,k-t+{\ell}-p)  \cr
& = & \frac{k^2}{6 \ell} \, c(k,k+\ell) \, \left(1+ O\left(1/\ell\right) + \Err(\ell, k) \right) \, ,
\label{eq:WAYS}
\een
where $\Err(\ell,k)$ satisfies for $1 \ll \ell \ll k$ 
\begin{equation}
\begin{array}\{{rrl}.
& \mathbf{(i)} & \Err(\ell,k) =     O\left( \sqrt{{\ell}^3/k}\right)\, , \, \mbox{if } \ell = o(k^{1/3}) \, \\ 
& \mathbf{(ii)} & \Err(\ell,k) = %
 O\left(\sqrt{\frac{\ell}{k}}\right)  + O\left( \frac{{\ell}^{1/16}}{k^{9/50}}\right) \, , \, %
 \mbox{if } \lim_{k \ten \infty} \frac{{\ell}^3}{k} \neq 0 \mbox{ and } %
 \ell \ll  k \, . 
\end{array}
\end{equation}
\end{theorem}

\noindent

Note here that our results differ from those in \cite{BCM92},
since we are interested in edges whose additions during
the random graph process, increase the complexity of
some connected components (whereas in \cite{BCM92} the results are more
general but all edges in a given connected component are considered with
the same probability).

Note also that Theorem \ref{THM_WAYS} will be used 
to compare the r.v.  $Y_n^{({\ell})}$ and $Z_n^{({\ell})}$.
We follow the probabilistic methods initiated by Janson and combine
them with the enumerative/analytic methods to study the moments of
the r.v. $X_n^{({\ell})}, Y_n^{({\ell})}$ and $Z_n^{({\ell})}$ described above,
 for values of ${\ell}$ and $n$ s.t.~ ${\ell}, n \ten \infty$ but ${\ell}=o(n)$.
More precisely, to obtain the results presented here,
methods of the probabilistic random graph process \Gt,
 studied in \cite{Ja93,Ja2000}, are combined with asymptotic enumeration
methods, developed by Wright in \cite{Wr77,Wr80} and by Bender, Canfield and McKay
 in \cite{BCM90,BCM92}.

\pl
\noindent $\mathbf{\bullet}$~ We turn on the expectations 
of the size and growth of components according to $\ell$ and find:
\begin{theorem} \label{SIZE}
Let $\VL$ be the number of
vertices that at some stage of the random graph process
belong to an ${\ell}$-component. 
As $n, \ell \ten \infty$, but ${\ell} = o(n)$, we have
\ben
\qE(\VL) \sim (12 {\ell})^{1/3} \, n^{2/3} \, .
\een
Let $X_n^{(\ell)}$ be the r.v. defined as the number of creations of $({\ell}+1)$-components
 during the evolution of the random graph and denote by
$Y_n^{({\ell})}$ (resp. $Z_n^{({\ell})}$) the number of  $({\ell}+1)$-components
created by the transition $\ell \rightarrow \ell+1$ (resp. 
$(\ell-p) \oplus p \rightarrow \ell+1, \, \ell\geq p \geq 0$) 
then as $n, \ell, \frac{n}{\ell} \rightarrow \infty$
\ben
\qE( X_n^{(\ell)}) \sim \qE(Y_n^{(\ell)}) \sim 1 \, \mbox{ and } %
\qE(Z_n^{(\ell)}) = O\l( \frac{1}{\ell} \r) \, .
\een
\end{theorem}

\pl
\noindent $\mathbf{\bullet}$~ We then 
obtain for the number of
$({\ell}+1)$-components, for $1 \ll {\ell} \ll n$,
created during the evolution of the graph:
\begin{theorem} \label{THM1}
Provided that the newly created $(k,k+{\ell}+1)$ component
satisfies ${\ell}=o(k^{1/3})$ then $Y_n^{(\ell)} \dist 1$
and $Z_n^{(\ell)} \dist 0$.
\end{theorem}
Note that in \cite[Section 16, Theorem 9]{JKLP93}, the authors obtained
the asymptotic probability that a random graph 
of a given configuration evolves to another configuration (see for instance
\cite[Section 16, Figure 1]{JKLP93}). Among other
results, they observed the evolution of complex components and 
proved that the probability that an evolving graph acquires
exactly $i \geq 1$ new complex components converges to
${p'}_i$ with ${p'}_1 \approx 0.87266$,
${p'}_2 \approx 0.12120$, ${p'}_3 \approx 0.00598$,
${p'}_4 \approx 0.00015$ (cf. \cite[Eq (27.15)]{JKLP93}).
In other words, the probability that an evolving graph
never has more than $4$ complex components is strictly greater
than $0.999998$. Theorem \ref{THM1} confirms this 
general tendency and we give here an alternative method,
 connecting the one from generating functions
initiated in \cite{Wr77} to those in \cite{Ja93}.

\subsection{\bf Outline of the paper}
This paper is organized as follows. The next section gives
the enumerative results of this paper (namely the proof of theorem \ref{THM_WAYS}).
In section 3, we compute the expectations  of the creations of $(\ell+1)$-component
as well as the expected number of vertices that ever belong to such components.
Section 4 offers the results about the moments of the random variables 
 $Y_n^{(\ell)}$ and $Z_n^{(\ell)}$. 
The limit distributions are obtained when studying the factorial moments
of these variables.


\section{ \bf Enumerating complex graphs with distinguished bridge}
As mentioned in paragraph \ref{OUR_RESULTS}, to investigate $(Z_n^{({\ell})})_m$, i.e.,
the number of ordered $m$-tuples of edges  added
between pairs of complex components to build an $({\ell}+1)$-component,
we will use tools from enumerative/analytic methods.

The enumeration of connected labelled graphs goes back to Cayley.
Denote by $T(z)$ the well-known
exponential generating function (EGF) of Cayley's
rooted trees \cite{Cayley}, we have
\ben T(z) = z\exp{(T(z))} = %
\sum_{n=1}^{\infty} \frac{n^{n-1} z^n}{n!} \, ,
\een
where the variable $z$ is associated to the labelled vertices.
(\EIS{A000169}\footnote{References to EIS correspond to
specific entries in \cite{Encyclopedia}.}).

Next, R\'enyi \cite{Ren59} found the EGF $W_{0}$ of unicyclic graphs.
\ben
W_0(z) = -\frac{1}{2} \ln{(1-T(z))} - %
\frac{T(z)}{2} - \frac{T(z)^2}{4} \, .
\label{eq:RENYI}
\een

More generally, Wright \cite{Wr77} found a recurrence formula
satisfied by the EGFs of ${\ell}$-components. Denote by $W_{\ell}(w,z)$ the bivariate EGF
of ${\ell}$-components where the variable $w$ marks the number of edges
and the variable $z$ the vertices. Thus, if $c(n,n+{\ell})$ is the number
of $(n,n+{\ell})$ connected graphs with $n$ vertices, we can write
\ben
W_{\ell}(w,z) = \sum_{n} c(n,n+{\ell}) w^{n+{\ell}} \frac{z^n}{n!} \,
\label{WRIGHT_EGFS}
\een
and Wright's recurrence formula \cite{Wr77} can be
stated as follow~:
\begin{eqnarray}
\V_w {W}_{{\ell}+1} = %
 w \Big( \frac{{\V_z}^2 - \V_z}{2} - \V_w \Big) {W}_{\ell}
 +  {w \over 2} \left(%
\sum_{p=-1}^{{\ell}+1} (\V_z {W}_{p}) (\V_z {W}_{{\ell}-p}) \right) \, ,
\label{FUNCTIONAL_WRIGHT}
\end{eqnarray}
where we denote by $\V_w$, resp. $\V_z$,
the differential operator $w \frac{\partial}{\partial w}$, resp. $z
\frac{\partial}{\partial z}$. Thus, the operator $\V_w$
corresponds to marking an edge present in a graph.
Similarly, $\V_z$ corresponds to marking a vertex.
The combinatorial pointing operator reflects the
distinction of an object among all the others.
For the use of pointing and marking, we refer to
\cite{GJ83} and  for general techniques
concerning graphical enumerations we refer to \cite{HP73}. All these  
EGFs are given  and explained in details in \cite{JKLP93}.
In terms of coefficients, (\ref{FUNCTIONAL_WRIGHT}) reads
\ben
& & (k+{\ell}+1)\, c(k,k+{\ell}+1) = %
  \left({k \choose 2} -k -{\ell}\right) \, c(k,k+{\ell}) \cr
& & \, \, \, \, +  \, \, \, \, \frac{1}{2} \sum_{t=1}^{k-1} \sum_{p=-1}^{{\ell}+1} %
{k \choose t} t(k-t) \, c(t,t+p) \, \,  c(k-t,k-t+{\ell}-p)  \, .
\label{eq:WRIGH_COEFFS}
\een
Starting with the differential equation (\ref{FUNCTIONAL_WRIGHT}),
Wright \cite{Wr77,Wr80} proved that each  $W_{\ell}$ can be written as~:
\begin{equation}
{W}_{\ell}(z) = \frac{b_{\ell}}{(1-T(z))^{3{\ell}}} - \frac{c_{\ell}}{(1-T(z))^{3{\ell}-1}} %
+ \sum_{ 2 \leq s \leq 3{\ell}-2} \frac{\omega_{{\ell},s}}{(1-T(z))^{s}} \, , %
\,\,\, ({\ell} \geq 1) \, ,
\label{EQ:TH_WRIGHT_FORM}
\end{equation}
where the coefficients $(b_{\ell})$ and $(c_{\ell})$
are rationals and more importantly, the summation is {\it finite}.
(Sequences for ${\ell}$-components are given by
\EIS{A061540} --- \EIS{A061544}~ for respectively $ \ell = 1, 2, \cdots, 5$).
The $(b_{\ell})_{\ell \geq 1}$ are called
the Wright's constants of first order (also called
Wright-Louchard-Tak\'acs constants, see  \cite{Sp97});
 $b_1=\frac{5}{24}$ and for $\ell \geq 1$, $b_{\ell}$ is defined recursively by
\begin{equation}
2 (\ell+1) b_{{\ell}+1} = 3{\ell}({\ell}+1)b_{\ell}+ 3 \sum_{p=1}^{l-1} t(\ell-p)b_p b_{l-p} \, .
\label{EQ:B_K}
\end{equation}
Note that the sequence $(c_{\ell})$ in (\ref{EQ:TH_WRIGHT_FORM})
 verifies also the following~:
\be
 2 (3{\ell}+2) c_{{\ell}+1} = 8 ({\ell}+1) b_{{\ell}+1} + 3{\ell} b_{\ell} + (3{\ell}+2)(3{\ell}-1) c_{\ell}
  + 6\sum_{p=1}^{l-1} p (3{\ell} - 3p -1) b_t c_{l-p} \, .
\label{EQ:C_K}
\ee
To study the asymptotic behavior of the coefficients
$c(k,k+{\ell})$, Wright \cite{Wr80} established that\footnote{
Remark that if $A(z)$ and $B(z)$ are two formal power series, the notation
$A(z) \preceq B(z)$ means that
$\forall n, \, \coeff{z^n}A(z) \leq \coeff{z^n}B(z)$.}:
\beq
    \frac{b_{\ell}}{(1-T(z))^{3{\ell}}} - \frac{c_{\ell}}{(1-T(z))^{3{\ell}-1}} %
  \preceq W_{\ell}(z) \preceq  \frac{b_{\ell}}{(1-T(z))^{3{\ell}}} \, ,
\label{eq:WRIGHT_INEQUALITIES}
\eeq
which we shall call {\it Wright's inequalities}.

We are interested in the number of creation  of $({\ell}+1)$-components.
In this Section,
we will study edges  added between a $p$-component and
a $(\ell-p)$-component, with $p\geq 0$.
Thus, we have to investigate the number of manners to build
a component with a distinguished cut edge.
The Theorem \ref{THM_WAYS} gives an estimate of
the number of such combinatorial structures.
It will be proved later since its proof involves
the decomposition of the Wright's EGFs
by means of inverse powers of $(1-T(z))$. In fact,
 Knuth and Pittel \cite{KP89} studied  combinatorially
and analytically the polynomial $t_n(y)$ defined  as follows
\begin{equation}
t_n(y) = n! \coeff{z^n} \frac{1}{\big( 1-T(z)\big)^y} \, ,
\label{EQ:TREE_POLYNOMIAL}
\end{equation}
which they call \textit{tree polynomial}.
The two authors observed that
the analysis of these polynomials can be used
 to study random graphs analytically as shown in \cite{FKP89,JKLP93}.
For our purpose, a very similar formula can be defined~:
\ben
t_{a,n}(y) = n! \coeff{z^n} \frac{T(z)^a}{\big( 1-T(z)\big)^y} \, .
\label{EQ:GENERAL_TREE_POLYNOMIAL}
\een

The lemma below is an application of the saddle point method
\cite{De Bruijn,FS+} to study the asymptotic behavior of the
coefficients $t_{a,n}(m)=n! \coeff{z^n} T(z)^a (1-T(z))^{-m(n)}$ as $m, \, n$
tend to infinity but $m\equiv m(n)=o(n)$.
\begin{lemma}  \label{TREE_POLYNOMIAL_INFINITY}
Let $\a \equiv \a(n)$ such that $\a \rightarrow 0$ as $n \ten \infty$
but $\a \, n \rightarrow \infty$, and
let $a$ and $\beta$ be fixed numbers. Then, $t_{a,n}(\a \, n+\beta)$
defined in (\ref{EQ:GENERAL_TREE_POLYNOMIAL}) satisfies
\ben
t_{a,n}(\a \, n+\beta) = \frac{n!}{2 \sqrt{\pi n}} %
\frac{\exp{ (n u_0) } (1-u_0)^{(1-\beta)}}{{u_0}^n (1-u_0)^{\a\,n}}  %
\left(1+O\Big(\sqrt{\a}\Big) + O\Big(\frac{1}{\a^{1/4} \,n^{1/4}} \Big)\right)
\label{EQ:TREE_POLYNOMIAL_INFINITY}
\een
where $u_0 = 1 + \frac{\a}{2} - \sqrt{\a(1+\frac{\a}{4})}  = 
1 - \sqrt{\a} + \frac{\a}{2} - \frac{\a^{3/2}}{8} + O(\a^2)$.
\end{lemma}

\begin{proof} 
Cauchy's integral formula gives (if we made the substitution $u=T(z)$ so that
$dz = e^{-u}(1-u)du$).
\ben
t_{a,n}(\a\, n+\beta) & = & n! \coeff{z^n}  %
\frac{T(z)^a}{\big( 1-T(z)\big)^{(\a\,n+\beta)}} \cr
 &=&   \frac{n!}{2 \pi i} \oint %
  \frac{T(z)^a} {\big(1-T(z) \big)^{\a\,n+\beta}} \frac{dz}{z^{n+1}}\cr %
 &= & \frac{n!}{2 \pi i} \oint %
 \frac{e^{nu}\, du}%
{(1-u)^{\a\,n+\beta-1} \, u^{n-a+1}  } \, .
\een
 The power $(\exp{(u)}/(1-u)^\a /u)^n$
suggests us to use the  saddle point method.
Let
\ben
h(u) = u - \ln(u) - \a \ln(1-u) \, .
\label{eq:HU}
\een
We then have
\ben
t_{a,n}(\a\,n+\beta) = \frac{n!}{2 \pi i} %
\oint \frac{(1-u)^{1-\beta}}{u^{1-a}} \exp(n h(u)) du \, .
\label{COL_TROIS}
\een
Investigating the roots of $h^{'}(u) = 0$, we find
two saddle points, at
\begin{eqnarray*}
u_0 = 1 + \frac{\a}{2} - \sqrt{\a(1+\frac{\a}{4})} \, \mbox{ and }\, 
u_1 = 1 + \frac{\a}{2} + \sqrt{\a(1+\frac{\a}{4})}\, .
\end{eqnarray*}
We remark that
\begin{eqnarray*}
h^{''}(u_0) = 2 + 3 \sqrt{\a} + O(\a) \,\mbox{ and } \,
h^{''}(u_1) = 2 - 3 \sqrt{\a} + O(\a)\,.
\end{eqnarray*}
The main point of the application of the saddle
point method here is that $h^{'}(u_0)=0$ and
$h^{''}(u_0) > 0$, hence $nh(u_0 \exp{(i\theta)})$
is approximately $nh(u_0) - n {u_0}^2 h^{''}(u_0) \frac{\theta^2}{2}$
in the vicinity of $\theta = 0$. Integrating (\ref{COL_TROIS}) around
a circle passing vertically through $u=u_0$ leads to
\ben
t_{a,n}(\a\,n+\beta) = \frac{n!}{2 \pi}
\int_{-\pi}^{\pi} %
u_0^a \, e^{i a \theta} \,(1-u_0 e^{i \theta})^{1-\beta} \,%
 \exp( n h(u_0 e^{i\theta}) ) d \theta \,
\label{COL_QUATRE}
\een
where
\begin{equation}
h(u_0 e^{i\theta}) = u_0 \cos \theta + i u_0 \sin \theta %
- \ln u_0 - i \theta - \a \ln (1-u_0 e^{i \theta}) \, \, .
\end{equation}
Let us check that the contribution away from
$] - \theta_0, \, \theta_0 [$ is bounded away by the
integrand at $\theta_0$.
Denote by ${\EuFrak{Re}}(z)$ the real part of $z$, we have
\begin{eqnarray}
f(\theta)  &=&  {\EuFrak{Re}}( h(u_0 e^{i \theta}))  \cr
  & = & u_0 \cos \theta - \ln u_0 - %
\a \ln ( |1-u_0e^{i\theta}|) \cr
 & =&  u_0 \cos \theta  - \ln u_0 - %
 \a \ln u_0 - \frac{\a}{2} %
\ln \big( 1+ \frac{1}{u_0^2} - \frac{2}{u_0} \cos \theta  \big) \, .
\end{eqnarray}
It comes
\begin{eqnarray}
f^{'}(\theta) &=& %
\frac{d}{d \theta}  \left(\EuFrak{Re}(h(u_0 e^{i\theta})) \right)\cr
&=& %
- u_0 \sin \theta - %
\frac{\frac{\a}{2}\big( \frac{2}{u_0} \sin \theta \big)^2}%
{\big(1 + \frac{1}{u_0^2} - \frac{2}{u_0} \cos \theta \big)}
\end{eqnarray}
and $f^{'}(\theta) = 0$ if $\theta = 0$. Also, $f(\theta)$ is
a symmetric function of $\theta$ and in
$\left[ -\pi, -\theta_0 \right] \cup \left[ \theta_0, \pi \right]$, for
a given $\theta_0$, $0  < \theta_0 < \pi$, it takes it maximum value for
$\theta = \theta_0$.

Since $|\exp( h(u))| = \exp( \EuFrak{Re}(h(u)) )$,
for a given $\theta_0$, $\theta_0 < \pi$,
when splitting
the integral in (\ref{COL_QUATRE}) into three parts, viz.
``$\int_{-\pi}^{-\theta_0} + \int_{-\theta_0}^{\theta_0}
 + \int_{\theta_0}^{\pi}$'', we know that it suffices to integrate
from $-\theta_0$ to $\theta_0$, for a convenient value of
$\theta_0$, because
the others can be bounded by the magnitude of the integrand
at $\theta_0$.

In fact, we have
\begin{eqnarray}
h(u_0 e^{i\theta}) & & = h(u_0)+ \frac{{u_0}^2 %
  (e^{i\theta}-1)^2}{2!}h^{''}(u_0) %
             +\frac{{u_0}^3 (e^{i\theta}-1)^3}{3!}h^{(3)}(u_0)
                   + \sum_{p \geq 4} %
\frac{{u_0}^p (e^{i\theta}-1)^p}{p!}h^{(p)}(u_0)   %
\cr
  & & = h(u_0) + \sum_{p\geq 2} \xi_p (e^{i\theta} -1)^p \quad ,
\label{ETOILE4}
\end{eqnarray}
 where $\xi_p = \frac{{u_0}^p}{p!} h^{(p)}(u_0)$.

The next computations are useful to estimate the error made when replacing $h(u_0 e^{i\theta})$
with an approximation.
 \noindent
For $p \geq 2$, we compute
$ h^{(p)}(u_0) = (p-1)! \Big( \frac{(-1)^p}{{u_0}^{p}} +
\frac{\a}{{(1-u_0)}^{p}}\Big)$, for $p \geq 2$ and for $\xi_p$, we have
\begin{eqnarray}
\xi_p & & =  \frac{ (-1)^p}{p} \Big(1 - \frac{\a \, {u_0}^p}%
                                  {(1-u_0)^p} \Big) \cr
         & & = \frac{ (-1)^p}{p} + %
         \frac{ (-1)^{p+1}}{p} \, %
 \frac{\a \, (1+\frac{\a \,}{2} - \sqrt{\a \,(1+\frac{\a \,}{4})})^p}{{\a \,}^{\frac{p}{2}} %
 (\sqrt{1+\frac{\a \,}{4}} - \frac{\sqrt{\a \,}}{2} )^p }  \, .
\label{ALPHA_P}
\end{eqnarray}
Thus, for $\a$ small enough and $p > 2$, 
we have 
\begin{equation}
| \xi_p | \leq \frac{2^p}{\a^{\frac{p}{2}-1}} \, ,  %
\, \, \, (\a \rightarrow 0, \, p>2) \,.
\end{equation}
On the other hand,
\begin{equation}
| e^{i\theta} - 1 | = \sqrt{ 2(1- \cos \theta)} < \theta %
\,, \,\, (\theta > 0)\, .
\end{equation}
Thus, the summation in (\ref{ETOILE4}) can be bounded for values of
$\theta$ and $\a$ such that $\theta \rightarrow 0$,
$\a \rightarrow 0$ but $\frac{\theta}{\sqrt{\a}} \rightarrow 0$ and we have
\ben
| \sum_{p \geq 4} \xi_p (e^{i\theta} - 1)^p | %
    \leq \sum_{p \geq 4} | \xi_p \theta^p |
    \leq  \a \, \sum_{p \geq 4}  \frac{2^p \theta^p}{\a^{\frac{p}{2}}} %
       =  O\Big( \frac{\theta^4}{\a} \Big) \, .
\een
It follows that for $\theta \rightarrow 0$,
$\a \rightarrow 0$ and $\frac{\theta}{\sqrt{\a}} \rightarrow 0$,
\begin{eqnarray}
h(u_0e^{i\theta}) & & = h(u_0) %
 - \frac{1}{2} \, \frac{u_0}{(1-u_0)^2} (1+\a -2u_0+{u_0}^2) \theta^2 \cr
& & + i \frac{u_0}{6(1-u_0)^3} \, %
(1+\a+(\a-3)u_0+3{u_0}^2 - {u_0}^3) \theta^3 %
+ O\Big(\frac{\theta^4}{\a}\Big)\, ,
\end{eqnarray}
where the term in the big-oh takes into account the terms from
$(e^{i\theta}-1)^2$ and $(e^{i\theta}-1)^3$ of (\ref{ETOILE4})
which we can neglect since
$$(e^{i\theta} -1)^2 = - \theta^2 - i \theta^3
+ O(\theta^4) \, \mbox{ and } \, 
(e^{i\theta}-1)^3 = -i\theta^3 + \frac{3}{2}\theta^4 +
i O(\theta^5)\,.$$
Let
\begin{eqnarray*}
\theta_0 = \frac{\a^{1/8}}{n^{3/8} \tau^{1/2}} \qquad \,\mbox{ with  } \, \qquad 
\tau = \frac{u_0(1+\a-2u_0+{u_0}^2)}{(1-u_0)^2} \, . 
\end{eqnarray*}
We can now use the magnitude of the integrand at $\theta_0$
to bound the resulting error.
Hence, we can verify our choice of $\theta_0$
\begin{eqnarray}
& & |u_0^{a} (1-u_0 e^{i\theta_0})^{(1-\beta)} %
 \left(\exp{( n h(u_0 e^{i \theta_0} ))}  - n h(u_0)\right) |  =   \cr
& &  u_0^{a}  |1-u_0 e^{i\theta_0} |^{(1-\beta)}  %
\exp \Big( - \frac{n}{2} \tau \, {\theta_0}^2  %
+ O\big( n \frac{{\theta_0}^4}{\a} \big) \Big) =
 O\Big( e^{-\frac{{\a^{1/4} \, n^{1/4}}}{2} } \Big)  \, .
\end{eqnarray}
To estimate $t_{a,n}(\a \,n+\beta)$, it proves convenient to compute the integral
\begin{equation}
 \int_{-\theta_0}^{\theta_0} u_0^{a} %
\exp{(ia\theta)} %
(1-u_0 e^{i \theta})^{(1-\beta)} %
\exp{(nh(u_0 e^{i\theta}))} d\theta \, .
\label{eq:J_N}
\end{equation}
If we make the substitution $\theta = \frac{t}{\sqrt{n \tau}}$,
we have (recall that $\theta_0 = \frac{\a^{1/8}}{n^{3/8} \tau^{1/2}} $)
\begin{equation}
\frac{u_0^{a}}{\sqrt{n \tau}} \int_{ - \a^{1/8} n^{1/8}}^{ \a^{1/8} n^{1/8}} %
\Big( 1 - u_0  e^{\frac{it}{\sqrt{n\tau}}}\Big)^{(1-\beta)} %
\exp\Big( ia \frac{t}{\sqrt{n\tau}} +  %
nh(u_0 e^{\frac{it}{\sqrt{n\tau}}}) \Big) dt \, .
\label{eq:J_N_1}
\end{equation}
Since 
$(1-u_0  e^{\frac{it}{\sqrt{n\tau}}})^{(1-\beta)} =(1-u_0)^{(1-\beta)} (1+O(t/\sqrt{n \a}))$,  
 the integral given in (\ref{eq:J_N}) becomes
\begin{eqnarray*}
\frac{1}{\sqrt{n \tau}} \, u_0^{a} \int_{- \a^{1/8} n^{1/8} }^{  \a^{1/8} n^{1/8} }
(1-u_0)^{(1-\beta)} &\exp&\Big(nh(u_0) -\frac{t^2}{2} %
+ i a \frac{t}{\sqrt{n \tau}} + %
 i f_3 \frac{t^3}{\sqrt{n \, \a}} + %
O\big(\frac{t^4}{n \, \a}\big) \Big)\\
&&\Big(1+O\big(\frac{t}{\sqrt{n \, \a}}\big) \Big) \,
dt
\end{eqnarray*}
where
$$f_3 = - \frac{\sqrt{\a}(1+\a+(\a-3)u_0+3u_0^2-u_0^3)}%
{\sqrt{u_0}(1+\a-2u_0+u_0^2)^{\frac{3}{2}}}= %
- \frac{\sqrt{2}}{12} -\frac{5}{48}\sqrt{\a} + O(\a)\, .$$
Using these approximations, we then obtain
\begin{eqnarray}
& &  u_0^{a} %
\frac{(1-u_0)^{(1-\beta)}}{\sqrt{n\tau}} e^{(nh(u_0))} \cr
 & &\, \,  \times   \left[ \, \int_{- \a^{1/8} n^{1/8}}^{ \a^{1/8} n^{1/8} } e^{- \frac{t^2}{2}}%
 \cos{\Big( f_3\frac{t^3}{\sqrt{n\a}} + a \frac{t}{\sqrt{n\tau}} \Big)}%
 \left( 1+ O\Big(\frac{t}{\sqrt{n\a}} \Big) + %
O\Big( \frac{t^4}{n\a}\Big)\right)\, dt \right] \, , 
\label{eq:32}
\end{eqnarray}
since the symmetry of the function leads to the cancellation of the terms
with the function $\sin$. Using, $u_0^{a} = 1+ O(\sqrt{\a})$,
$\cos{(x)} = 1 + O(x^2)$ and $\exp{(O(x))} = 1+O(x)$ when $x=O(1)$ 
 in (\ref{eq:32}), we find
\begin{eqnarray}
& & \frac{(1-u_0)^{(1-\beta)}}{\sqrt{n\tau}} e^{(nh(u_0))} \cr
 & &\, \, \times  \left[  \int_{- \a^{1/8} n^{1/8}}^{ \a^{1/8} n^{1/8} } e^{- \frac{t^2}{2}}%
 \left( 1+ O\Big(\frac{1}{\a^{1/4} \, n^{1/4}} \Big) + %
O\Big( \frac{\a^{1/4}}{n^{3/4}} \Big)    \right)\, dt \right] %
\left( 1+ O(\sqrt{\a}) \right) \cr
&=& \frac{(1-u_0)^{(1-\beta)}}{\sqrt{n\tau}} e^{(nh(u_0))} \cr
  & &\, \, \times  \left[ \, \int_{-\infty}^{\infty} e^{- \frac{t^2}{2}}%
 \left( 1+ 
O\Big( \frac{1}{\a^{1/4}\, n^{1/4}} \Big)    \right)\, dt %
\, \, + \, \, O\Big( e^{- \frac{\a^{1/4} \, n^{1/4}}{2} } \Big)  \right] %
\left( 1+ O(\sqrt{\a}) \right) \cr
&=& \frac{\sqrt{2\pi} (1-u_0)^{(1-\beta)} e^{(nh(u_0))} } {\sqrt{n\tau}} %
\left( 1 \, +  \, O\Big(\sqrt{\a}\Big) + O\Big(\frac{1}{\a^{1/4}\, n^{1/4}} \Big) %
\,  + \,  O\Big( e^{- \frac{\a^{1/4} \, n^{1/4}}{2} } \Big)  \right) \cr
&=& \sqrt{\frac{\pi}{n}}(1-u_0)^{(1-\beta)} e^{(nh(u_0))} %
\left( 1 \, +  \, O\Big(\sqrt{\a}\Big) + O\Big( \frac{1}{\a^{1/4}\, n^{1/4}} \Big) \right) \,.\cr %
\label{COL_THE_END}
\end{eqnarray}
\end{proof}

Now, we are ready to prove Theorem \ref{THM_WAYS}. The proof is divided into
two parts according to the range of excess. In the first part $\mathbf{(i)}$, we consider
connected  graphs with $k$ vertices and $k+o(k^{1/3})$ edges and the methods in use
are due to Wright \cite{Wr80}. In the second part $\mathbf{(ii)}$, we will consider
excesses with wider range and the methods are those of Bender-Canfield-McKay \cite{BCM90,BCM92}.

\begin{proof}[Proof of Theorem \ref{THM_WAYS}: Part $\mathbf{(i)}$.]
In term of EGFs, $c'(k,k+{\ell}+1)$ represents the coefficient
\ben
c'(k,k+{\ell}+1) = {k!\over 2} \coeff{z^k} %
 \sum_{p=0}^{\ell} \B( \V_z W_p(z) \B) \B( \V_z W_{l-p}(z) \B) \, .
\label{eq:POINTING_AGAIN}
\een
Applying Wright's inequalities , i.e. (\ref{eq:WRIGHT_INEQUALITIES}),
 in (\ref{eq:POINTING_AGAIN}) yields
\ben
L_{\ell}(z) \preceq  \sum_{k} c'(k,k+{\ell}+1) \frac{z^k}{k!} \preceq R_{\ell}(z) \, ,
\, \, (\ell>0) \, ,
\label{eq:APPLY_WRIGHT_INEQ}
\een
where
\ben
R_{\ell}(z) = \frac{9}{2}\sum_{p=1}^{l-1} %
\frac{ p(\ell-p)b_p b_{l-p} T(z)^2}{\b(1-T(z) \b)^{3{\ell}+4}} %
+ \frac{3 {\ell} b_{\ell} T(z)^5}{2 (1-T(z))^{3{\ell}+4}}\,
\label{eq:RIGHT}
\een
and
\ben
L_{\ell}(z) = R_{\ell}(z) - %
\left( \sum_{p=1}^{l-1} %
\frac{3(3p-1)(\ell-p) b_{l-p}c_p T(z)^2}{(1-T(z))^{3{\ell}+3}} %
+ \frac{(3{\ell}-1)c_{\ell} T(z)^5}{2(1-T(z))^{3{\ell}+3} }\right) \, .
\label{eq:LEFT}
\een
(We used $\V_z T(z) = T(z)/(1-T(z))$.)
Our aim is then to show that the difference
between the coefficients of the right and left parts
of (\ref{eq:APPLY_WRIGHT_INEQ}),
viz. $k! \coeff{z^k} (R_{\ell}(z) - L_{\ell}(z))$ can be neglected in comparison
to $k! \coeff{z^k} R_{\ell}(z)$ for $\ell=o(k^{1/3})$.
For this purpose, we   use lemma  \ref{TREE_POLYNOMIAL_INFINITY},
and
the fact that  $b_{\ell} = (\frac{3}{2})^{\ell} (\ell-1)! d_{\ell}$ with $(d_{\ell})$  an
increasing sequence tending to $1\over {2\pi}$ (cf. \cite[eq. (1.4)]{Wr80},
\cite{BCM90}).

More precisely, lemma  \ref{TREE_POLYNOMIAL_INFINITY} tells
us that in $R_{\ell}(z)$, the coefficients of $z^k$ of $T(z)^2/(1-T(z))^{3{\ell}+4}$ and
$T(z)^5/(1-T(z))^{3{\ell}+4}$ in (\ref{eq:RIGHT}) have
the same order of magnitude for $\ell= o(k^{1/3})$.
Next, using the definition of Wright's coefficients (\ref{EQ:B_K}),
we find
\ben
 \frac{9}{2}\sum_{p=1}^{\ell-1}  p(\ell-p)b_p b_{\ell-p} %
+ {3 \over 2} \ell b_{\ell} = %
3({\ell}+1)(b_{{\ell}+1} - {3 \over 2} \ell b_{\ell}) + {3 \over 2} \ell b_{\ell}\, .
\een
We then have
$$b_{{\ell}+1} - {3 \over 2} \ell b_{\ell} =
  ({3 \over 2})^{{\ell}+1} {\ell}! (d_{{\ell}+1} - d_{\ell})$$
where we used $b_{\ell} = (3/2)^{\ell} (\ell-1)! d_{\ell}$ as studied
in \cite[eq. (1.4)]{Wr80} and in \cite{JKLP93}.
>From the proof given by Meertens in \cite[lemma 3.4]{BCM90},
we have $ 0 < d_{{\ell}+1} - d_{\ell} = O(1/{\ell}^2).$
So,
\ben
 \frac{9}{2}\sum_{p=1}^{\ell-1}  p(\ell-p) \, b_p \, b_{\ell-p} %
+ {3 \over 2} \ell \, b_{\ell} = \B({3 \over 2}\B)^{{\ell}+1} \, %
{\ell}! \, d_{\ell} \B(1+O(1/{\ell}) \B) \, .
\een
On the other hand, the definition (\ref{EQ:C_K}) of the sequence
$(c_{\ell})$ tells us that the summation in (\ref{eq:LEFT})
satisfies $$\sum 3(3p-1)(\ell-p) b_{\ell - p}c_p = O(\ell c_{\ell})$$and we
know from \cite{Wr80} that $c_{\ell} = O(\ell b_{\ell})$.
Finally, lemma \ref{TREE_POLYNOMIAL_INFINITY} suggests
us  to find values of $\ell \equiv \ell(k)$ for which
the coefficients of the difference $R_{\ell} - L_{\ell}$ satisfy
$$\coeff{z^k}\b(R_{\ell}(z) - L_{\ell}(z) \b) \ll \coeff{z^k} \b( R_{\ell}(z) \b)\,.$$
It comes $\ell = o(k^{1/3})$ which is the same
range as in \cite{Wr80} and in \cite{Thesis} for connected graphs without
prefixed (forbidden) configurations, the error terms
being of order $O(1/{\ell})+O(\sqrt{{\ell}^3/k})$. After a bit of algebra, we find
(replacing $\a = 3{\ell}/k$ in the saddle point $u_0$)
\ben
\frac{3}{2} \, \ell \, b_{\ell} \, t_{5,3{\ell}+4} & = & {3 \over 2} \, \ell \, b_{\ell} %
 \frac{k^{k+3/2{\ell}+3/2}} {\sqrt{2} (3{\ell})^{3\ell/2+3/2}} \exp{(3{\ell}/2)} %
 \B( 1+O(\sqrt{{\ell}^3/k}) \B) \cr
& = & {1 \over \sqrt{48 \, \pi} \, \ell} \,  %
   \B( {e\over 12l} \B)^{{\ell}/2} k^{k+3{\ell}/2+3/2} %
 \B( 1+O(1/{\ell}) + O(\sqrt{{\ell}^3/k}) \B) \, ,
\een
which completes the proof of Theorem \ref{THM_WAYS} part $\mathbf{(i)}$.
\end{proof}

Wright showed that the EGFs of all multicyclic components
can be expressed in terms of the EGF of Cayley. In order to count the
number of ways to label a complex component, one can repeatedly
prune it by deleting recursively any vertex of degree $1$.
The graph obtained after removing all vertices of degree $1$ is called
a {\it smooth graph}. The process of removing recursively all
vertices of degree $1$ is called \textit{smoothing} or \textit{pruning} process \cite{Wr78}.
\begin{rem}
 Theorem \ref{THM_WAYS} tells us that asymptotically
 almost all $({\ell}+1)$-components whose situation after smoothing
contains a cut edge are built by linking a unicyclic
component to another complex component. In fact, 
(\ref{eq:WAYS}) reflects simply
\ben
c'(k,k+{\ell}+1) \sim k! \coeff{z^k}\b(\V_z W_0(z)\b)\b( \V_z W_{\ell}(z) \b)%
 \, , \qquad  1 \ll \ell \ll k^{1/3} \, .
\een
\end{rem}

Using the same
technics involved in the  proof of Theorem \ref{THM_WAYS} part $\mathbf{(i)}$, 
we obtain a generalization:
\begin{corollary} \label{COROLLARY_WAYS}
Denote by $c^{r}(k,k+{\ell}+1)$ the number of manners to build an
$({\ell}+1)$-component of order $k$ with a distinguished edge between
a $p$-component and an $(\ell-p)$-component, with $p \geq r \geq 0$.
Then, as $k,{\ell} \ten \infty$, $\ell=o(k^{1/3})$ and for fixed values
of $r$, we have
\ben
c^{r}(k,k+{\ell}+1) \sim k! %
 \coeff{z^k}\b(\V_z W_r(z)\b)\b( \V_z W_{\ell-r}(z) \b)%
 \, , \qquad  1 \ll \ell \ll k^{1/3} \, .
\label{eq:WAYS_COR}
\een
\end{corollary}

\begin{rem} \label{UPPER_BOUND_CR}
Observe that the value of $h(u_0)$ with $h$ given by (\ref{eq:HU}) and $u_0$ given
in Lemma \ref{TREE_POLYNOMIAL_INFINITY} satisfies
\ben
h(u_0) = 1+ \left( \frac{1}{2}+ \frac{1}{2} %
\,\ln  \left( \frac{1}{\rho} \right)  \right) \, \rho + \frac{1}{3} \,
{\rho}^{3/2}-{\frac {1}{120}}\,{\rho}^{5/2}+ O \left( {\rho}^{3}  \right) %
\, , \quad (\rho \rightarrow 0) \, .
\label{eq:APPROX_HU}
\een
Thus, for the range $\ell=o(k)$, it is also possible
to obtain an upper-bound of $c^{r}(k,k+{\ell}+1)$ (for any fixed integer $r \geq 0$) 
by means of the same methods as above 
 and we then get
\ben
c^{r}(k,k+{\ell}+1)  & \leq & \frac{1}{\sqrt{48 \pi} \, \ell} \, %
\B(e/12 \ell\B)^{{\ell}/2}\,  k^{k+3{\ell}/2+3/2}  \, %
\exp{\left( \frac{3^{1/2} \, \ell^{\, 3/2} }{k^{1/2}} \right)} \, \cr
 & \times & \l( 1 + O\B(\frac{1}{\ell^{1/4}}\B) %
      + O\B(\sqrt{\frac{\ell}{k}}\B) \r) \, .
\label{eq:UPPER_BOUND}
\een
\end{rem}

\begin{proof}[Proof of Theorem \ref{THM_WAYS}: Part $\mathbf{(ii)}$.]
The second part of the proof is entirely different and is based
 upon the results in \cite{BCM90,BCM92}. We start comparing $c'(k,k+\ell)$
with $( {k \choose 2} - k - \ell + 1)c(k,k+\ell -1)$ by means of (\ref{FUNCTIONAL_WRIGHT}) 
and (\ref{eq:WRIGH_COEFFS}).
Using the definition of $c'(k,k+\ell)$, 
 \cite[Lemma 4.1 and (4.12)]{BCM90} and denoting $q=k+\ell$, we have
\ben
\frac{c'(k,q)}{q \, c(k,q)} &= & %
\frac{\sum_{t=1}^{k-1} {k \choose t} t\, c(t,t) \, (k-t) \, c(k-t,q-t-1)} %
{q { {k \choose 2} \choose q} \exp{\B(k \varphi(x)+a(x)\B)} \B(1+b(k,\ell)\B) } \, \, + \, \, \cr
 &  & %
\frac{1}{2}\sum_{s=1}^{\ell-2} \sum_{t=1}^{k-1}%
 \frac{ {k \choose t} t\, c(t,t+s) \, (k-t) \, c(k-t,q-t-s-1)} %
{q { {k \choose 2} \choose q} \exp{\B(k \varphi(x)+a(x)\B)} \B(1+b(k,\ell)\B) }  \, ,
\label{SPLIT_DEUX}
\een
where $x=q/k=1+\ell/k$, $\varphi$ and $a$ are, respectively, given in \cite[(1.12)]{BCM90} and
\cite[(1.17)]{BCM90}, and the error term $b(\cdot,\,\cdot)$ is given in
\cite[(1.20a) and (1.20b), Theorem 2]{BCM90}.
Thus, we have
\ben
\frac{\B( 1 + b(k,\ell) \B) \, c'(k,q)}{q \, c(k,q)} = S_0 + S \, ,
\een
where (again) $S_0$ and $S$ are defined in \cite[equations (4.2) and (4.3)]{BCM90}, i.e.,
 the first and the second summations in the equation (\ref{SPLIT_DEUX}) above.
Hence, the quantity of interest is given by
\ben
&  & \frac{c'(k,q)}{\l( {k \choose 2} - q + 1 \r) c(k,q-1)} = %
\frac{S_0+S}{(1+b(k,\ell))} \times \frac{q c(k,q)}{\l( {k \choose 2} -q+1 \r) c(k,q-1)}  \cr
 & &  =  \frac{(S_0 + S)}{(1+b(k,\ell))} \times %
\frac{ q { {k \choose 2} \choose q} %
\exp{\l( k \varphi(x) +  a(x) \r)}\, \l(1+b(k,\ell)\r)} %
 { \l( {k \choose 2} -q+1 \r) { {k \choose 2} \choose q-1} %
\exp{\l( k \varphi(x- {1 \over k}) +  a(x -{1 \over k}) \r)}\, \l(1+b(k,\ell-1)\r)} \cr
& & = \frac{S_0 + S}{ 1   +  b(k, \ell-1)} \times
\exp{ \l( \varphi'(x) - {1 \over 2k} \varphi''(x) %
+ {1 \over 6k^2} \varphi^{(3)}(x-\theta_\varphi)%
+ {a'(x) \over k} - {a''(x-\theta_a) \over 2k^2} \r) } %
\, \, , \cr
& & \,
\een
 where $\theta_\varphi$ and $\theta_a$ are in $(0, \, {1 \over k})$.
Taking into account the bounds given in \cite[Lemma 3.1]{BCM90}, 
 viz.,  ${\varphi''(x)  \over k} = O(1/\ell)$, 
 ${\varphi^{(3)}(x-\theta_\varphi) \over k^2} = O(1/k^2)$,
${a'(x) \over k} = O(1/\sqrt{\ell k})$ and ${a''(x-\theta_a) \over k^2} = O(k^{-1/2} \ell^{-3/2})$,
we get
\ben
& & \frac{c'(k,q)}{\l( {k \choose 2} - q + 1 \r) c(k,q-1)} = %
\qquad \qquad  \qquad \qquad e^{\varphi'(x)}  %
\qquad \times \qquad \l(S_0 + S\r) \qquad \times \cr
& &  \l(1 + O \l({1 \over \ell}\r) %
                      + O \l( {1 \over k^2} \r) %
                      + O \l({1 \over \ell^{1/2} k^{1/2}}\r)%
                      + O \l({1 \over \ell^{3/2} k^{1/2}}\r)%
                      + O \l({\ell^{1/16} \over k^{9/50}}\r) \r) \, \cr
& &  \qquad \qquad  \qquad \qquad \qquad \quad =  \qquad \sqrt{k \over 3 \ell} %
\qquad \times \qquad \l(S_0 + S\r) \qquad \times \cr
& &  \l ( 1 +  O \l({1 \over \ell}\r) %
                      + O \l({1 \over \ell^{1/2} k^{1/2}}\r) %
		+ O \l( \ell \over k \r) %
                      + O \l({\ell^{1/16} \over k^{9/50}}\r) \r)  \, . \cr        
& & \, 
\label{eq:PENULTIMATE}         
\een
Now, we can use the approximations of $S_0$ and $S$ given 
respectively by \cite[equation (4.6c)]{BCM90} and by \cite[equation (4.6d)]{BCM90} 
to get (after a bit of algebra)
\ben
S_0+S = \frac{1}{\sqrt{3 \ell k}} \l( 1+ O\l( \sqrt{ {\ell \over k}} \r) \r) \, .
\label{eq:ULTIMATE}         
\een
The combination of (\ref{eq:PENULTIMATE}) and (\ref{eq:ULTIMATE}) completes the proof.
\end{proof}

\section{ \bf Expectations of transitions and size of $\mathbf{\ell}$-component }

When adding an edge in a randomly growing graph, there is a possibility
that it joins two vertices of the same component, increasing its excess
by $1$ (transition $\ell \rightarrow {\ell}+1$).

Consider an ${\ell}$-component 
with $k$ vertices. Let $\alpha(\ell;k)$ be the expected number of times that a new edge is added
to an ${\ell}$-component of order $k$ (with both ends of the edge in the component). 
When a new edge is added to an ${\ell}$-component of order $k$,
there are ${n \choose k} c(k,k+{\ell})$ manners to choose an ${\ell}$-component and 
${k \choose 2} - k - \ell$ ways to choose the new edge. Furthermore,
the probability that such possible component is one of $\mathbb{G}(n,t)$
 is $t^{k+{\ell}} (1-t)^{(n-k)k + {k \choose 2} - k - \ell}$ and with
the conditional probability $\frac{dt}{(1-t)}$ that a given edge is
added during the interval $(t,t+dt)$ and not earlier, integrating
over all times, we obtain (see also \cite{Ja2000})
\begin{equation}
\alpha(\ell;k) = {n \choose k} c(k,k+{\ell}) \l(\frac{k^2 - 3k - 2\ell}{2} \r) %
\int_{0}^{1} t^{k+{\ell}} (1-t)^{(n-k)k + {k \choose 2} - k - \ell -1} dt
\label{eq:FORMULE_INTEGRAL}
\end{equation}
which evaluation leads to 
\begin{equation}
\alpha(\ell;k) = {(n)}_k \frac{(k+{\ell})!}{k!} c(k,k+{\ell}) %
 \frac{(k^2 -3k - 2\ell)}{2} %
\frac{(nk - k^2/2 - 3k/2 - \ell - 1)!}{(nk - k^2/2 - k/2)!} \, .
\label{eq:LEMMA11}
\end{equation}

\pl

For the second type of transition $(\ell-p) \oplus p \rightarrow \ell+1$ 
(with $0 \leq p \leq \ell$), let $\beta(\ell-p, p; k_1, k-k_1)$ be the expected
number of times an edge is added between an $(\ell-p)$-component of size $k_1$ and
a $p$-component of size $k-k_1$. 
Since there are $k_1 (k-k_1)$ manners to join two fixed $(\ell - p)$-component
and $p$ component of order $k_1$, respectively $k-k_1$, 
instead of (\ref{eq:FORMULE_INTEGRAL}), we have
\ben
\beta(\ell - p, p; k_1,k-k_1) &= &  {n \choose k} {k \choose k_1} %
  k_1 c(k_1, k_1 + \ell - p)  \, \, (k-k_1) c(k-k_1, k-k_1 + p)  \cr
& &  \qquad \qquad \qquad %
\times \int_{0}^{1} t^{k+{\ell}} (1-t)^{(n-k)k + {k \choose 2} - k - \ell -1} dt \, .
\een
When summing over $p$ and $k_1$, we then obtain
\ben
 & &  {n \choose k} \, \, \underbrace{ %
\l( \frac{1}{2} \, \sum_{p=0}^{\ell} \sum_{k_1=1}^{k-1}{k \choose k_1} %
\, \,  k_1 c(k_1, k_1 + \ell - p)  \, \, (k-k_1) c(k-k_1, k-k_1 + p) \r)} \cr
& \times&  \int_{0}^{1} t^{k+{\ell}} (1-t)^{(n-k)k + {k \choose 2} - k - \ell -1} dt \, ,
\een
and we recognize that the double-summation represents exactly the 
coefficient $c'(k,k+\ell+1)$ defined in Theorem \ref{THM_WAYS}. Therefore, 
 the second kind of transition
can be deduced using the first one simply by introducing a factor
$O(\frac{1}{\ell})$ as indicated by (\ref{eq:WAYS}). 

\pl

Before proving Theorem \ref{SIZE}, we need several  lemmas which are
given in the next paragraph.

\subsection{Technical lemmas}
We have the following result which gives bounds of $\alpha(\ell;k)$:

\begin{lemma} \label{LEMMA1} 
As $\ell, k, n \ten \infty$ but
$\ell = o(k)$, we have 
\ben
\alpha(\ell;k) &\leq& \frac{1}{4} \, \sqrt{ \frac{3}{\pi}} \, %
\l( \frac{e}{12 \ell}\r)^{\ell/2} \, \frac{k^{3\ell/2+1/2}}{n^{\ell+1}} \, 
\exp{\l(-\frac{k^3}{24n^2} + \frac{k^4}{n^3} %
+ \frac{2 \ell k}{n} + \frac{\ell^2}{2 k} + \frac{3^{1/2} \, \ell^{3/2}}{k^{1/2}} \r)} \, \cr
& \times & \left( 1 + O\l( {k \over n} \r)  + O\l( \sqrt{\frac{\ell}{k}} \r) %
+ O\l( \frac{1}{\ell^{1/4}} \r)  \right) \, , \qquad \qquad (k \leq n) \, ,
\label{EQ:UPPER_ALPHA}
\een
and 
\ben
\alpha(\ell;k) & \geq &  \frac{1}{4} \, \sqrt{ \frac{3}{\pi}} \, %
\l( \frac{e}{12 \ell}\r)^{\ell/2} \, \frac{k^{3\ell/2+1/2}}{n^{\ell+1}} \, 
\exp{\l(-\frac{k^3}{24n^2} - \frac{k^4}{n^3} + \frac{\ell k}{2 n} -\frac{\ell^3}{6 k^2}\r)} \, \cr
& \times & \left( 1 + O\l( {k \over n} \r)  + O\l( \sqrt{\frac{\ell}{k}} \r) %
+ O\l( \frac{1}{k} \r) \right) \,  , %
\qquad \qquad (k \leq \frac{n}{2}) \, .
\label{EQ:LOWER_ALPHA}
\een
\end{lemma}

\noindent \textbf{Proof.} 
The proof given here are based on the works of \auteur{Janson} in
\cite{Ja93,Ja2000}. However, the main difference comes from the
fact that our parameter, representing the excess of the 
sparse components ${\ell}$, is no more fixed as in \cite{Ja2000}.
Allowing $\ell$ to grow smoothly with $n$ introduces new difficulties.

For $1 \leq k \leq n$ and $\ell = o(n)$, the value of the
integral in (\ref{eq:FORMULE_INTEGRAL}) is
\ben
\frac{(nk - k^2/2 - 3k/2 - \ell - 1)!}{(nk - k^2/2 - k/2)!} & = & %
k^{-k -{\ell} -1} (n - k/2)^{-k - \ell - 1} \cr
 & \times & \l(1 + O\l(\frac{k}{n}\r) + %
 O\l(\frac{\ell}{n}\r) + %
O\l( \frac{\ell^2}{k \, n} \r) \r) \, .
\label{eq:LEMMA13}
\een
We have,
\ben
\frac{{(n)}_k}{(n-k/2)^k} 
 & = & \exp\B( \sum_{i=1}^{k-1} \ln{\l(1-{i \over n}\r)} - k \ln{\l( 1- {k \over 2 n}\r)} \B) \cr
 & \leq  & %
\exp\B( \sum_{i=1}^{k-1} \ln{\l(1-{i \over n}\r)} + {k^2 \over 2 n}  %
                  + {k^3 \over 8 n^2} + {k^4 \over n^3} \B) \cr
 & \leq & \exp\B( - {k^3 \over 24 n^2} + {k^4 \over n^3}  \B) \, %
        \B( 1+ O\l( {k \over n} \r) \B) \, ,
\label{eq:LEMMA14}
\een
and assuming that $1 \leq k \leq {n \over 2}$ we find the following lower-bound
\ben
\frac{{(n)}_k}{(n-k/2)^k} \geq 
 \exp\B( - {k^3 \over 24 n^2} - {k^4 \over n^3}  \B) %
 \,     \B(1      + O\l( {k \over n} \r) \B) \, , \qquad (1 \leq k \leq {n \over 2}) \, .
\label{eq:LEMMA14BIS}
\een
(We used $\ln{(1-x)} \geq -x -x^2/2 - 4x^3\, , x \in \coeff{0, \, 1/2}$.) \\
\noindent
Obviously ${k \choose 2} - k - \ell \leq \frac{k^2}{2}$ and
\begin{equation}
{k \choose 2} - k - \ell = \frac{k^2}{2} \l( 1 + O(1/k) + O(\ell/k^2) \r) \, .
\label{eq:LEMMA15}
\end{equation}
Thus, combining  (\ref{eq:LEMMA13}), (\ref{eq:LEMMA14}), (\ref{eq:LEMMA14BIS})
and (\ref{eq:LEMMA15}) in (\ref{eq:LEMMA11}), we infer that
\ben
\alpha(\ell;k) &\leq& \frac{1}{2} \frac{(k+{\ell})!}{k!} %
\frac{c(k,k+{\ell})}{(n-k/2)^{\ell+1} k^{k+\ell-1}} %
\exp\B(- \frac{k^3}{24 n^2} + \frac{k^4}{n^3} \B) \, \cr
& \times &  %
\B( 1+ O\l(\frac{k}{n}\r) + O\l(\frac{\ell^2}{kn} \r) + O\l(\frac{\ell}{n}\r) \B) \, %
\qquad ( 1 \leq k \leq n)  \qquad \qquad \mbox{ and } \cr
\alpha(\ell;k) & \geq & \frac{1}{2} \frac{(k+{\ell})!}{k!} %
\frac{c(k,k+{\ell})}{(n-k/2)^{\ell+1} k^{k+\ell-1}} %
\exp\B(- \frac{k^3}{24 n^2} - \frac{k^4}{n^3} \B) \, \cr
& \times &  %
\B( 1+ O\l(\frac{k}{n}\r) + O\l(\frac{\ell^2}{kn} \r)  %
+ O\l(\frac{1}{k}\r)  + O\l( \frac{\ell}{k^2}\r) + O\l(\frac{\ell}{n}\r) \B) \, %
\qquad ( 1 \leq  k \leq {n \over 2})  \, .
\label{eq:LEMMA16}
\een
Also, we simply get (using $-2x \leq \ln(1-x) \leq -x\, , x \in \coeff{0, \, 1/2}$)
\ben
\exp{\l(- \frac{2 \ell k}{n} \r)} %
\leq \frac{(n-k/2)^{{\ell}+1}}{n^{{\ell}+1}} \leq %
\exp{\l( - \frac{\ell k}{2 n} \r)} \, .
\label{eq:LEMMA18}
\een
Taylor expansions lead to
\ben
 \frac{\ell^2}{2 k} - \frac{\ell^3}{6 k^2} +  O\l( \frac{\ell}{k} \r) \, 
\leq \ln{\l( \frac{(k+\ell)!}{k^{\ell} k!} \r)} %
\leq \frac{\ell^2}{2 k} + O\l( \frac{\ell}{k} \r) \, .
\label{eq:LEMMA17}
\een
Now, by Wright's inequality and by means of (\ref{EQ:TREE_POLYNOMIAL_INFINITY}),
we can get an upper-bound of the quantity $c(k,k+\ell)$ above. After a bit of algebra,
we then have
\ben
c(k,k+\ell) & \leq &  \frac{1}{2} \, \sqrt{\frac{3}{\pi}} %
 \l( \frac{e}{12 \ell}\r)^{\frac{\ell}{2}}%
k^{k+ 3/2 \ell - 1/2} %
\, \exp{\left( \frac{3^{1/2} \ell^{3/2}}{k^{1/2}} \right)} \, \cr
& \times & \l(1 + O\l( \frac{1}{\ell^{1/4}}\r) %
+ O\l( \sqrt{\frac{\ell}{k}} \r)  \r) %
\, .
\een
Using the result  \cite[Theorem 3]{PITTEL-WORMALD}~ 
(whenever $\ell/k \ten 0$ but $(k+\ell) \exp{(-2(k+\ell)/k)} \ten \infty$),
one can get a lower-bound
of the same quantity, viz.
\ben
c(k,k+\ell) & \geq & \frac{1}{2} \, \sqrt{\frac{3}{\pi}} %
 \l( \frac{e}{12 \ell}\r)^{\frac{\ell}{2}} \, 
k^{k+ 3/2 \ell - 1/2} 
\exp{\l( - \frac{\ell^2}{2 \, k} \r)} \, 
 \l( 1  
+ O\l( \sqrt{\frac{\ell}{k}} \r)  \r) \, .
\een
Combining the above inequalities, we find the bounds of the quantity defined
by $\alpha(\ell;k)$.
\ENDPROOF

Since $\qE(Y_n^{({\ell})}) = \sum_{1 \leq k \leq n} \alpha(\ell;k)$,
the bounds given in lemma \ref{LEMMA1} 
 suggest us to consider the asymptotic behavior of sums of the form
\beq
\sum_{k} k^a %
\exp{ \left( -\frac{k^3}{24n^2} + c_1 \frac{k^4}{n^3} + 
 c_2 \frac{2\ell k}{n} + c_3 \frac{\ell^2}{2 k} %
+ c_4 \frac{\ell^3}{k^2} + c_5 \frac{\ell^{3/2}}{k^{1/2}} \right) }
\label{eq:FORM_SUM}
\eeq
where $a= \frac{3{\ell}+1}{2}$, $\ell \equiv \ell(n)$,  as $n \rightarrow \infty$
and the $c_i, \, 1 \leq i \leq 5$ are absolute constants.

Now, our plan is to show 
\[
\qE(Y_n^{({\ell})}) = \sum_{k=1}^{n} \alpha(\ell;k) %
 \sim   \sum_{k=\omega(n)}^{n/2} \alpha(\ell;k) \, 
\]
where 
\[
\omega(n) =  \frac{\ell}{10} \, \ln{\l(\frac{n}{\ell}\r)} \, .
\]
In the second part of the summation, the values of $k$ satisfy
$\ell \ll \omega(n) \leq k \leq n/2$. So, we can use
the bounds given in lemma \ref{LEMMA1}. Therefore, we  have to prove that
the sums $\sum_{k=1}^{\omega(n)-1}\alpha(\ell;k)$ and
$\sum_{k=n/2+1}^{n}\alpha(\ell;k)$ can be neglected.
In these directions, we have the following lemma:
\begin{lemma} \label{LEMMA1BIS}
As $1 \ll \ell \ll n$, set $\omega(n) = \frac{\ell}{10} \, \ln{\l(\frac{n}{\ell}\r)}$.
We have,
\ben
\sum_{k=1}^{\omega(n)} \alpha(\ell;k) = o\l( \frac{1}{\sqrt{n}} \r) \, .
\label{SQRTY}
\een
\end{lemma}

\noindent \textbf{Proof.} 
There is a constant $A$ s.t. $c(k,k+\ell) \leq (A/\ell)^{\ell/2} k^{k+3\ell/2-1/2}$
for every $k$ and $\ell$ (see for instance \cite{Bollobas}). Using this and
with similar bounds to those given during the proof of lemma \ref{LEMMA1}, we
successively  get (for $k \leq \omega(n)$ and $\ell = o(n)$):
\beq
 \alpha(\ell;k)  \leq   {(n)}_k \frac{(k+{\ell})!}{k!} \, \l(\frac{A}{\ell}\r)^{\ell/2}
 \, {k^{k+3\ell/2+3/2}} \, 
\frac{(nk - k^2/2 - 3k/2 - \ell - 1)!}{(nk - k^2/2 - k/2)!} 
\eeq
and since $\frac{(k+\ell)!}{k!} \leq \ell^{\ell} \exp{(2 k)}$
\beq
\alpha(\ell;k) \leq   {(n)}_k \exp{(2 k)} \l(A \ell\r)^{\ell/2} \,  \, {k^{k+3\ell/2+3/2}} \, 
\frac{(nk - k^2/2 - 3k/2 - \ell - 1)!}{(nk - k^2/2 - k/2)!} \, .
\eeq
In the considered ranges, we have
\beq
\frac{(nk - k^2/2 - 3k/2 - \ell - 1)!}{(nk - k^2/2 - k/2)!}  \leq %
2\, \frac{\exp{\l( \frac{4 \ell^2}{k n} \r)} } {k^{k+\ell+1} (n-k/2)^{k+\ell+1}}  \, .
\eeq
and (since $k \leq n$)
\beq
\frac{1}{(n-k/2)^{\ell+1}} = \frac{1}{n^{\ell+1}} \, %
\exp{\B( -(\ell+1) \ln(1-k/2n) \B)} \leq \frac{ \exp{\l(  \frac{(\ell+1) k}{n} \r)}}{n^{\ell+1}} %
\leq \frac{\exp{\l( k \r)}}{n^{\ell+1}} \,.
\eeq
Combining the above inequalities with (\ref{eq:LEMMA14}), we get
\ben
\alpha(\ell;k) & \leq & 2\, \frac{ \l( A \ell \r)^{\ell/2} }{n^{\ell+1}} \, k^{\ell/2+1/2} \,
\exp{\l( -\frac{k^3}{24 n^2} + \frac{k^4}{n^3} + 3k + \frac{4 \ell^2}{k n} \r)} \cr
& \leq & 2\, \frac{ \l( A \ell \r)^{\ell/2} }{n^{\ell+1}} \, k^{\ell/2+1/2} \,
\exp{\l( 5 k + 4 \ell \r)} \, .
\een
Therefore,
\ben
\sum_{k=1}^{\omega(n)} \alpha(\ell;k) %
& \leq & %
2\,  \frac{(A e^{8} \ell)^{\ell/2}}{n^{\ell+1}} 
\sum_{k=1}^{\omega(n)} k^{\ell/2+1/2} \exp{(5k)} %
 \leq  %
2\,  \frac{(A e^{8} \ell)^{\ell/2}}{n^{\ell+1}} 
\, \exp{(5 \, \omega(n) )} \, \sum_{k=1}^{\omega(n)} k^{\ell/2+1/2} \cr
& \leq & O\l( \frac{1}{\sqrt{n}} \, %
\frac{\ln{\l(\frac{n}{\ell}\r)}^{3/2}}{\sqrt{\l(\frac{n}{\ell}\r)} } %
\, \l( \frac{\tilde{A} \ln{\l(\frac{n}{\ell}\r)}} {\l( \frac{n}{\ell} \r) }  \r)^{\ell/2}
 \r) \, .
\label{eq:77}
\een
($\tilde{A}$ is some constant.)
\ENDPROOF


\noindent
For summation of the form described in (\ref{eq:FORM_SUM}), we have the
following approximation
\begin{lemma} \label{LEMMA2} 
Let $c>0$ and $c_1, c_2, c_3, c_4$ be fixed constants. If $\ell, n \ten \infty$
but $n \gg \ell$ and 
$a \equiv a(\ell) = \Theta(\ell)$ then we have
\ben
\sum_{k=1}^{n/c} \exp{\l(\phi(k,n,\ell)\r)} & \equaldef & \sum_{k=1}^{n/c} k^a %
\exp \l( - \frac{k^3}{24 n^2} + c_1 \frac{k^4}{n^3} + 
 c_2 \frac{\ell k}{n} %
+ c_3 \frac{\ell^2}{k} + c_4 \frac{\ell^3}{k^2} + c_5 \frac{\ell^{3/2}}{k^{1/2}} \r) \cr
& \sim & 2^{a+1} 3^{(a-2)/3}  \Gamma \l(\frac{a+1}{3} \r) n^{2(a+1)/3} \, .
\label{eq:LEMMA2}
\een
\end{lemma}

\noindent \textbf{Proof.}  
We have
\ben
\sum_{k=1}^{n/c} \exp{\l(\phi(k,n,\ell)\r)} \sim %
\int_{1}^{n/c} \exp{\l(\phi(t,n,\ell)\r)} dt \, .
\label{eq:SUM_INT}
\een
If we denote by 
$I_n$ the integral, we have after substituting $t$ for $2 n^{2/3}e^z$:
\ben
I_n & = &  2^{a+1} n^{\frac{2(a+1)}{3}} %
\int_{-\frac{2}{3}\ln{n} - \ln{2}}^{\frac{1}{3} \ln n - \ln{(2c)}} %
\exp(H(z)) dz \, .
\label{eq:I_n}
\een
where 
\ben
& & \, \cr
& & H(z)  =  \left( a+1 \right) z- \frac{e^{3\,z}}{3} +16\,{\frac {c_{{1}}{e^{4\,z}}}{{n}^{1/3}}}
+{\frac {c_{{2}}{\it \ell}\,{e^{-z}}}{2 {n}^{5/3}}}+
{\frac {c_{{3}}{{\it \ell}}^{2}{e^{-z}}}{2 {n}^{2/3}}}+%
{\frac {c_{{4}}{{\it \ell}}^{3}{e^{-2\,z}}}{4 {n}^{4/3}}} +
{\frac {c_{{5}}{{\it \ell}}^{3/2}{e^{-z/2}}}{2^{1/2} {n}^{1/3}}} \, .
\label{eq:H}
\een
Also, we have
\ben
H'(z) & = & a+1-{e^{3\,z}}+64\,{\frac {c_{{1}}{e^{4\,z}}}{{n}^{1/3}}}-{
\frac {c_{{2}}{\it \ell}\,{e^{-z}}}{2 {n}^{5/3}}}-{\frac {c_{{3}}{{
\it \ell}}^{2}{e^{-z}}}{2 {n}^{2/3}}}-{\frac {c_{{4}}{{\it \ell}}^{3}
{e^{-2\,z}}}{2 {n}^{4/3}}} -
{\frac {c_{{5}}{{\it \ell}}^{3/2}{e^{-z/2}}}{2^{3/2} {n}^{1/3}}}
\label{eq:H1}
\een
and more generally (for $q>1$)
\ben
H^{(q)}(z) & = & - 3^{q-1} e^{3\, z} + 4^{q+2} \,{\frac {c_{{1}}{e^{4\,z}}}{{n}^{1/3}}} %
+ (-1)^{q} \frac {c_{{2}}{\it \ell}\,{e^{-z}}}{2 \, {n}^{5/3}} + %
 (-1)^{q} {\frac {c_{{3}}{{\it \ell}}^{2}{e^{-z}}}{2\, {n}^{2/3}}} \cr 
&+ & %
 (-2)^{q-2} {\frac {c_{{4}}{{\it \ell}}^{3}{e^{-2\,z}}}{{n}^{4/3}}} +
{\l( - \frac{1}{2}\r)^q \frac {c_{{5}}{{\it \ell}}^{3/2}{e^{-z/2}}}{2^{1/2} {n}^{1/3}}} \, .
\label{eq:H2}
\een
Let $z_0$ be the solution of $H^{'}(z) = 0$. 
By hypothesis, $a=\Theta(\ell)$ so that $a$ is large. Therefore, $z_0$ is located near
$\frac{1}{3} \ln (a+1)$. We can proceed by
an iterative method (see \cite[Chapter 2]{De Bruijn}) to get a full
asymptotic expansion of $z_0$.
 For our present purpose the first few terms of such expansion suffice. If we
let $x_0 = \exp{(z_0)}$, solving  $H^{'}(z_0) = 0$ we obtain
\ben
{x_0}^3 %
& = & (a+1) + O\l( \frac{\ell^{4/3}}{n^{1/3}} \r) \, .
\label{eq:APPROXZ_0}
\een
We also have
\ben
H(z_0) & = & \frac{a+1}{3}\ln{(a+1)} -  \frac{a+1}{3} %
+ O\l(\frac{{\ell}^{4/3}}{n^{1/3}}\r)
\label{eq:hZ_0}
\een
and 
\ben
H^{''}(z_0) &= & -  3(a+1) +  O\l(\frac{{\ell}^{4/3}}{n^{1/3}}\r) \, .
\een
That is $H^{''}(z_0) <0$. At this stage, we can consider 
$\exp \l( H^{''}(z_0) \frac{(z-z_0)^2}{2} \r)$ as the main
factor of the integrand. We refer here to the book of \auteur{De Bruijn}
 \cite[\S 4.4 and \S 6.8]{De Bruijn} for more discussions about asymptotic 
estimates on integrals of the forms 
``$\int {x^{a}\, e^{\tiny\mbox{Polynomial}\normalsize(x)}}$''
  and we infer that
\begin{equation}
\int_{-\frac{2}{3}\ln{n} - \ln{2}}^{\frac{1}{3} \ln n - \ln{(2c)}} %
\exp(H(z)) dz \sim %
\sqrt{ - \frac{2 \pi}{H^{''}(z_0)}} %
\exp \l(H(z_0)\r) \, .
\label{eq:BEFORE-STIRLING}
\end{equation}
Using the Stirling formula for Gamma function, i.e., 
$\Gamma(t+1) \sim \sqrt{2 \pi t} \frac{t^t}{e^t}$
and since  $z_0 \sim \frac{1}{3} \ln (a+1)$,
$H(z_0) \sim \frac{(a+1)}{3} (\ln (a+1)-1) $ and $H^{''}(z_0) \sim -3(a+1)$,
we can see that (\ref{eq:BEFORE-STIRLING}) leads to (\ref{eq:LEMMA2}) which is
similar to the formula already obtained by \auteur{Janson} in \cite{Ja2000}.
\ENDPROOF

\subsection{ \bf Proof of Theorem \ref{SIZE} }
Using lemmas \ref{LEMMA1}, \ref{LEMMA1BIS}, \ref{LEMMA2} 
(namely with $a=\frac{3\ell+1}{2}$) and theorem \ref{THM_WAYS},
 after nice cancellations, we find the results announced in the theorem.

Now, let us describe briefly how to proceed. Using the upper-bound
given in (\ref{EQ:UPPER_ALPHA}) valid for $1 \ll \ell k \leq n$ and
lemma \ref{LEMMA1BIS}, we obtain
\ben
\qE(Y_n^{(\ell)}) &\leq& o\l(\frac{1}{\sqrt{n}}\r) + %
\frac{1}{4} \, \sqrt{3}{\pi} \, \l( \frac{e}{12 \ell}\r)^{\ell/2} 
\, 2^{3\ell/2 + 3/2} \, 3^{3\ell/2 - 1/2} \Gamma\l( \frac{\ell}{2} + \frac{1}{2} \r) %
\l( 1 +o(1)\r) \, \cr
& \leq & 1+o(1) \, .
\een
Next, using the lower-bound (\ref{EQ:LOWER_ALPHA}) and summing only for
$\omega(n) \ll k \ll n/2$, we get (using lemma \ref{LEMMA1BIS}) 
$\qE(Y_n^{(\ell)}) \geq 1 + o(1)$.
We find $\qE(Z_n^{(\ell)}) \sim \frac{1}{3\ell}$ by Theorem \ref{THM_WAYS} and also
\ben
\qE(\VL) &  \sim & \l( \sum_{k=\omega(n)}^{n} k \alpha(\ell-1;k) \r)\l( 1 + O\l(\frac{1}{\ell}\r) \r)
 \sim 12^{1/3} \ell^{1/3} n^{2/3} \, .
\een

\section{ \bf Higher moments }
To simplify computations, we consider in the rest of the paper 
only  $(k,k+o(k^{1/3}))$ connected graphs.

\subsection{Adding edges to an ${\ell}$-component: higher moments}
As already said, proves given here follow (humbly) the works of Janson in
\cite{Ja93,Ja2000} but in our work
we allow the parameter $\ell$ to grow relatively with $n$.
Turning to higher moments, we observe that $\qE(Y_n^{({\ell})})_m$ is
the number of $m$-tuples of edges added
to a $i-$th ${\ell}$-component of order $k_i$ during the evolution
of the random graph process.

There are ${n \choose {k_1 \ldots k_m}} \prod _i c(k_i,k_i+{\ell})$
manners to choose an ${\ell}$-component having respectively
$k_1, \ldots, k_m$ vertices. There are
$\prod _i\left({k_i \choose 2} - k_i - \ell \right)$
ways to choose the new edge.
 Furthermore,
the probability that such possible component is one of
\Gt is
$$
 \prod _i t_i^{k_i+{\ell}} (1-t_i)^{(n-\sum k_j)k_i %
+ {k_i \choose 2} - k_i - \ell} \prod _{i <j} %
\left( 1- t_i \vee t_j \right)^{k_ik_j}
$$
and with
the conditional probability $\frac{dt_i}{(1-t_i)}$ that a given edge is
added during the interval $(t_i,t_i+dt_i)$ and not earlier, integrating
over all times, i.e. $t_i \in \left[0, \, 1\right]$ and summing
over $k_i$, we obtain
\ben
\qE(Y_n^{({\ell})})_m &=&\sum_{k_1=1}^{n} \ldots \sum_{k_m=1}^{n} %
\int _0^1\ldots \int _0^1(n)_\kp %
\prod _i \frac{c(k_i,k_i+{\ell})}{k_i!} %
\left({k_i \choose 2} - k_i - \ell  \right) \cr
 &&t_i^{k_i+{\ell}} (1-t_i)^{(n-k_i)k_j + {k_i \choose 2} - k_i - \ell -1}%
\prod _{i <j} \left( 1- t_i \vee t_j \right)^{k_ik_j} dt_1 \ldots dt_m
\label{eq:hm}
\een
where $\kp=\sum_i k_i$. We remark here that
$$c(k_i,k_i+{\ell}) = 0 \mbox{ for } k_i = 1, 2, \cdots \lceil (3+\sqrt{9+8 \ell })/2 \rceil -1\,.$$
Rewriting the integrand in (\ref{eq:hm}) as a function
of $k_i$ and $t_i$,  viz.
$$\varphi_n(k_i,t_i) \equiv \varphi_n(k_1, \cdots, k_m, t_1, \cdots, t_m)\, ,$$
with $\varphi_n(k_i,t_i) = 0$ if $\exists j \in \left[1,m\right]$~ s.t.~
$k_j \leq  \lceil (3+\sqrt{9+8 \ell})/2 \rceil -1$ or $k_j > n$ or
$t_j \notin (0,1)$ and
substituting $k_i=\lceil x_in^{2/3}\rceil$ and $t_i=n^{-1}+u_in^{-4/3}$,
we have
\ben
{\qE({Y_n}^{({\ell})})}_m &=& \int_{0}^{n^{1/3}} \cdots \int_{0}^{n^{1/3}} %
  \int_{-n^{1/3}}^{n^{4/3}-n^{1/3}} \cdots   %
  \int_{-n^{1/3}}^{n^{4/3}-n^{1/3}} \cr
&  & \, \, \, \times \, \, \,
 \varphi_n\left(\lceil x_i n^{2/3} \rceil , %
\frac{1}{n}+\frac{u_i}{n^{4/3}}\right) %
\frac{ du_i \cdots du_m \, dx_i \cdots dx_m}{n^{2m/3}} \cr
& = & \int_{0}^{\infty} \cdots \int_{0}^{\infty} %
 \int_{-\infty}^{\infty} \cdots \int_{-\infty}^{\infty} %
\Psi_n^{(m)}\left( x_i,u_i\right)  du_i \cdots du_m \, dx_i \cdots dx_m \, ,
\label{eq:hm2}
\een
where
\ben
\Psi_n^{(m)}\left(x_i,u_i \right) \equiv  %
\Psi_n^{(m)}\left(x_1, \cdots, x_m, \,u_1 , \cdots, u_m \right) =
\frac{\varphi_n \left( \lceil x_i n^{2/3} \rceil , %
\frac{1}{n}+\frac{u_i}{n^{4/3}} \right) }{n^{2m/3}} \, .
\een
We shall now investigate the integrand in (\ref{eq:hm}).
For this purpose, we consider each term of the products
in this integrand and
we assume that $x_i n^{2/3}$ are integers. In the following,
for each factor, we use the substitutions $k_i=x_i n^{2/3}$
and $t_i = n^{-1} + u_i n^{-4/3}$ as done above. We then have
(denoting $\xp = \sum x_i$)
\ben
{(n)}_{\kp} = n^{\kp} %
\exp{ \left( - \frac {{\xp}^2}{2} n^{1/3} - \frac{\xp^3}{6} %
+ O\B(\frac{\xp}{n^{1/3}}(1+\xp^3) \B) \right)} \, ,
\label{eq:Prod1}
\een
\ben
{k_i \choose 2} - k_i - \ell
 & = & \frac{{k_i}^2}{2} %
\B( 1+ O\B(\frac{1}{k_i}\B) \B) \cr
     & = & \frac{x_i^2}{2} n^{4/3} %
\B( 1+ O\B(\frac{1}{x_i n^{2/3}}\B) \B) \, .
\label{eq:Prod2}
\een
Using Stirling's formula and asymptotic formulae for $c(k_i,k_i+{\ell})$
(see for instance \cite{BCM92, Wr80}), it
yields
\ben
\frac{c(k_i,k_i+{\ell})}{k_i!}
& = & %
\sqrt{\frac{3}{2}} d \, \exp{\B( \frac{\ell}{2} + k_i \B)} %
{k_i}^{\frac{3}{2}\ell -1} \frac{1}{(12\, \ell )^{{\ell}/2}} %
\left( 1+ O\B( \frac{1}{\ell}\B) %
+ O\B( \frac{1}{k_i} \B) +  %
O\B(\frac{{\ell}^{3/2}}{k_i^{1/2}} \B)  \right) \cr
& =& \sqrt{\frac{3}{2}} d \, \exp{\B( \frac{\ell}{2} + x_i n^{2/3} \B)} %
{x_i}^{\frac{3}{2} \ell -1} n^{\ell-\frac{2}{3}} \frac{1}{(12\, \ell)^{{\ell}/2}} \cr
&  & \, \, \times \, \, \left( 1+ O\B( \frac{1}{\ell}\B) %
+ O\B( \frac{1}{x_i n^{2/3}} \B) +  %
O\B(\frac{{\ell}^{3/2}}{{x_i}^{1/2} n^{1/3}} \B)  \right) \,.
\label{eq:Prod3}
\een
(We use formulae $c(k_i,k_i+{\ell})$ for $l = o(k_i^{1/3})$ and
$d = \frac{1}{2 \pi}$ as described in \cite{BCM92}.)
Also, after the same substitutions
\ben
{t_i}^{k_i + l}
 & = & \frac{1}{n^{k_i+{\ell}}} %
   \B(1+\frac{u_i}{n^{1/3}}\B)^{k_i+{\ell}} \cr
 & = & \frac{1}{n^{\ell} n^{x_i n^{2/3}}} %
\exp\left( \frac{\ell u_i}{n^{1/3}} - \frac{\ell u_i}{2n^{2/3}} + %
x_i u_i n^{1/3} - \frac{x_i u_i^2}{2} %
+ O\B(\frac{\ell u_i^3}{n} \B) %
+ O\B(\frac{x_i u_i}{n^{1/3}} \B) \right) \, ,
\label{eq:Prod4}
\een
and
\ben
(1- t_i \vee t_j)^{k_i k_j}
& = & %
 \exp\left( -k_i k_j(t_i \vee t_j) %
+\frac{1}{2} k_i k_j(t_i \vee t_j)^2 %
+ O(k_i k_j)(t_i \vee t_j)^{3}\right) \cr
 &= & \exp\left(-x_i x_j n^{1/3} - x_i x_j(u_i \vee u_j) + %
O\B(\frac{x_i x_j}{n^{2/3}} \B)  \right) \, .
\label{eq:Prod5}
\een
Finally,
\ben
(1-t_i)^{(n-\kp)k_i + {k_i \choose 2} - k_i - \ell - 1} & = & %
\exp\B( %
-x_i n^{2/3} - (u_i + \xp ) x_i n^{1/3} + %
 u_i x_i \xp  \cr
& & \,\,- %
\frac{x_i^2}{2} \left(n^{1/3}+ u_i\right)+ \frac{3}{2} \frac{x_i}{n^{1/3}} \left( 1+ \frac{u_i }{n^{1/3}}\right )\cr
	&& \,\, +
\frac{\ell}{n} + \frac{\ell u_i}{n^{4/3}} + \frac{1}{n}+ %
\frac{u_i}{n^{4/3}} + O\B(\frac{x_i}{n^{1/3}} \B) %
\B) \, .
\label{eq:Prod6}
\een
Using the equations (\ref{eq:Prod1}) -- (\ref{eq:Prod6}) given above, the integrand
in (\ref{eq:hm2}) reads
\ben
\Psi_n^{(m)}(x_i,u_i) =  A_m \exp{(B_m)} (1+\varepsilon) \, .
\label{eq:AVANT_LEBESGUE}
\een
A bit of algebra gives $A_m$ and $B_m$
\ben
A_m = \left(\sqrt{\frac{3}{2}}\right)^m d^m \prod_{i=1}^{m} %
\frac{ x_i^{3/2{\ell}+1}}{2} \, ,
\een
\ben
& & \cr
& & B_m = \frac{ml}{2}\B(1- \ln{(12 \ell )}\B) - \frac{\xp^3}{6} %
 -\sum_{i=1}^{m} \frac{x_i u_i^2}{2} + \xp %
\sum_{i=1}^{m} x_i u_i - \frac{1}{2} \sum_{ 1 \leq i, \, j \leq m} %
x_i x_j (u_i \vee u_j) \, . \cr
& &
\een
The $\varepsilon$ in (\ref{eq:AVANT_LEBESGUE}) regroups all
the big-Ohs produced by (\ref{eq:Prod1}) -- (\ref{eq:Prod6}).
In particular, if $(x_i)$, $(u_i)$ and $(1/x_i)$ are fixed,
as $n \ten \infty$, we have
\ben
\Psi_n^{(m)}(x_i,u_i) =  A_m \exp{(B_m)} (1+o(1)) \, .
\een
So, if $x_i>0$, $u_i \in (-\infty,\infty)$ fixed, without
restricting each $x_i n^{2/3}$ to be an integer, we get
\ben
& & \Psi_n^{(m)}(x_i,u_i)  \ten  \left(\sqrt{\frac{3}{2}}\right)^m d^m %
\frac{\exp{\left(\frac{m\ell}{2} \right)}}{\left(12 \ell \right)^{\frac{m\ell}{2}}} %
\left(\prod_{i=1}^{m} \frac{ x_i^{3/2{\ell}+1}}{2}\right) \cr
& & \, \, \, \times \, \, \, %
\exp{\left(  - \frac{\xp^3}{6} %
 -\sum_{i=1}^{m} \frac{x_i u_i^2}{2} + \xp %
\sum_{i=1}^{m} x_i u_i - \frac{1}{2} \sum_{ 1 \leq i, \, j \leq m} %
x_i x_j (u_i \vee u_j)\right) } \,
\label{eq:LEBESGUE1}
\een
as $n \ten \infty$.
Next, we   use the estimate
\[
\prod_{i<j} (1-t_i \vee t_j)^{k_i k_j} %
\leq \prod_{i \neq j} (1-t_i)^{k_i k_j/2} \, ,
\]
to state that, there is a constant $C_1$ 
such that
\ben
\Psi_n^{(m)} \leq C_1 {(n)_{\kp}} \prod_i %
\frac{c(k_i,k_i+{\ell})}{k_i!} %
\left(\frac{k_i^2-3k_i}{2} -{\ell} -1 \right){t_i}^{k_i} %
(1-t_i)^{k_i(n-2-\kp/2)} \, .
\een
Then, using the bounds given in \cite[eq.~ (2.12) -- (2.18)]{Ja93}
with (\ref{eq:Prod1}) -- (\ref{eq:Prod6}), we get
(the $C_i$ below are constants) 
\ben
\Psi_n^{(m)} & \leq  & g_m(x_i,u_i) = C_2 \exp{(-\delta \xp^3)} %
\prod_i x_i^{3{\ell}/2+1} \exp{(-\delta x_i u_i^2)} \cr
& & + \, \, \, %
C_3 \exp{(-\delta \xp^3)} %
\prod_i x_i^{3{\ell}/2+1} \exp{(-\delta x_i u_i)} \cr
& & + \, \, \, %
C_4 \exp{(-\delta \xp^3)} \prod_i \frac{1}{(1+u_i^2)} %
\, ,
\label{eq:TRICKS_JANSON}
\een
valid for all $n, x_i, u_i$. Since
$\int_{0}^{\infty} \cdots \int_{0}^{\infty}
\int_{-\infty}^{\infty} \cdots \int_{-\infty}^{\infty} g_m(x_i,u_i)
dx_1 \cdots dx_m du_1 \cdots du_m < \infty$,
(\ref{eq:hm2}), (\ref{eq:LEBESGUE1}) and the use of Lebesgue
dominated convergence yield
\ben
\qE( Y_n^{({\ell})})_m \sim \left(\sqrt{\frac{3}{8}}  %
\frac{d \, \exp{(\frac{\ell}{2})}}{(12\ell)^{\frac{\ell}{2}}} \right)^m a_m^{({\ell})} \, ,
\label{eq:ENDLEBESGUE1}
\een
where
\ben
a_m^{({\ell})} & = & \int_{0}^{\infty} \cdots \int_{0}^{\infty} %
\int_{-\infty}^{\infty} \cdots \int_{-\infty}^{\infty} %
\left( \prod_{i=1}^{m}  x_i^{3{\ell}/2+1}\right) \cr
& &\, \, \times \, \, \exp{\Big( %
-\frac{1}{6} \xp^3 - \frac{1}{2} \sum_{i=1}^{m} x_i u_i^2 + \xp %
\sum_{i=1}^{m} x_i u_i\Big)} \cr
& & \, \, \times \, \, \exp{\Big(- \frac{1}{2} \sum_{ 1 \leq i, \, j \leq m} %
x_i x_j (u_i \vee u_j)\Big) } dx_1 \cdots dx_m du_1 \cdots du_m \, \cr
& \leq &  \int_{0}^{\infty} \cdots \int_{0}^{\infty} %
\left( \prod_{i=1}^{m}  x_i^{3{\ell}/2+1}\right) \, %
 \exp{\l(- \frac{1}{24} \l(\sum_{i=1}^{m} x_i\r)^3  \r)} %
dx_1 \cdots dx_m \cr
& \leq &  \int_{0}^{\infty} \cdots \int_{0}^{\infty} %
\left( \prod_{i=1}^{m}  x_i^{3{\ell}/2+1}\right) \, %
 \exp{\l(- \frac{1}{24} \sum_{i=1}^{m} x_i^3  \r)} %
dx_1 \cdots dx_m \cr
&\leq & %
\l( \frac{4}{3} \,{2}^{\ell}{6}^{\ell/2}{3}^{2/3}\Gamma  \left( \ell/2 +2/3 \right) \r)^m \, .
\label{eq:AM}
\een
Using this latter inequality with (\ref{eq:ENDLEBESGUE1}), we get that
\ben
\qE(Y_n^{(\ell)})_m \rightarrow 0  \, \qquad (m > 0, \quad n, \, \ell \ten \infty) \, .
\een

\subsection{ \bf Joining two complex components: higher moments}
We observe that $(Z_n^{({\ell})})_m$ is the number of
$m$-tuples of edges  added between
a $p$-component and, resp., a $(\ell-p)$-component of order $k_i$
and, resp., $k_j$. By Theorem \ref{THM_WAYS}, we find
\ben
 & & \cr
 & & c'(k_i,k_i+{\ell}+1) = %
{1 \over 3{\ell}}  \B(k_i^2/2  - 3k_i/2 -{\ell}\B)\, c(k_i,k_i+{\ell})%
\B(1 + O(1/{\ell}) + O({\ell}^{3/2}/k_i^{1/2}) \B) \, \cr
 & & 
\een
which means that we can obtain expressions for $\qE(Z_n^{({\ell})})_m$ by simply
introducing a factor $1/3{\ell}$ in (\ref{eq:hm}).
Therefore, 
\ben
\qE( Z_n^{({\ell})})_m \ten 0, \qquad (m > 0, \quad n, \, \ell \ten \infty) \, .
\label{eq:ENDLEBESGUE_SIMILARITY}
\een

\section{ \bf Conclusion}
In this paper, we have studied the growths of complexity of
connected components in an evolving graph. We have shown, using
a combination of the methods from \cite{Ja93} and the
theory of generating functions, how one can quantify
asymptotically  properties of such components growths.
Amongst other things, we study complex components
that increase their complexity by receiving new edges and/or by merging
other complex components. As ${\ell} \ten \infty$,
our results show that whenever the second case occurs,
 almost all times, it is a unicyclic component that is
swallowed by the considered ${\ell}$-component.
Our other result states that as $1 \ll \ell \ll n$,
the expected number of vertices that ever belong to
an $\ell$-component is about $(12 \ell)^{1/3} \, n^{2/3}$.

\bibliographystyle{plain}

\end{document}